\documentclass[letterpaper,twocolumn,10pt]{article}
\usepackage{usenix-2020-09}
\usepackage[available]{usenixbadges}

\usepackage{graphicx}
\usepackage{hyperref}
\usepackage{multirow}
\usepackage[normalem]{ulem}
\useunder{\uline}{\ul}{}
\usepackage[utf8]{inputenc}
\usepackage[ruled, lined, longend, linesnumbered]{algorithm2e}
\usepackage{amsmath,multicol}

\usepackage{amssymb,subcaption,textcomp,comment}
\usepackage[normalem]{ulem}  %
\usepackage{wrapfig}
\usepackage{appendix}

\usepackage{booktabs}
\usepackage{multirow}
\usepackage{hhline}
\usepackage{colortbl}
\usepackage{tcolorbox}

\usepackage{array}
\newcolumntype{L}[1]{>{\raggedright\let\newline\\\arraybackslash\hspace{0pt}}m{#1}}
\newcolumntype{C}[1]{>{\centering\let\newline\\\arraybackslash\hspace{0pt}}m{#1}}
\newcolumntype{R}[1]{>{\raggedleft\let\newline\\\arraybackslash\hspace{0pt}}m{#1}}

\newcommand{\soccluster}{SoC Cluster\xspace}
\newcommand{\socclusters}{SoC Clusters\xspace}

\usepackage[framemethod=tikz]{mdframed}
\newcommand{\summary}[1]{
\vspace{2.5mm}
\begin{mdframed}[linecolor=gray,roundcorner=12pt,backgroundcolor=gray!15,linewidth=3pt,innerleftmargin=2pt, leftmargin=0cm,rightmargin=0cm,topline=false,bottomline=false,rightline = false]
#1
\end{mdframed}
\vspace{1mm}
}

\begin{document}
\date{}

	\title{More is Different: Prototyping and Analyzing a New Form of Edge Server with Massive Mobile SoCs}

	\author{
		{\rm Li Zhang$^1$, Zhe Fu$^2$, Boqing Shi$^1$, Xiang Li$^1$, Rujin Lai$^3$, Chenyang Yang$^3$}\\
		{\rm Ao Zhou$^1$, Xiao Ma$^1$, Shangguang Wang$^1$, Mengwei Xu$^1$}\\
		$^1$Beijing University of Posts and Telecommunications\\
		$^2$Tsinghua University, $^3$vclusters\\
		}

	\maketitle

	\begin{abstract}

Huge energy consumption poses a significant challenge for edge clouds.
In response to this, we introduce a new type of edge server, namely \soccluster, that orchestrates multiple low-power mobile system-on-chips (SoCs) through an on-chip network.
For the first time, we have developed a concrete \soccluster consisting of 60 Qualcomm Snapdragon 865 SoCs housed in a 2U rack, which has been successfully commercialized and extensively deployed in edge clouds.
Cloud gaming emerges as the principal workload on these deployed \socclusters, owing to the compatibility between mobile SoCs and native mobile games.

In this study, we aim to demystify whether the \soccluster can efficiently serve more generalized, typical edge workloads.
Therefore, we developed a benchmark suite that employs state-of-the-art libraries for two critical edge workloads, i.e., video transcoding and deep learning inference.
This suite evaluates throughput, latency, power consumption, and other application-specific metrics like video quality.
Following this, we conducted a thorough measurement study and directly compared the \soccluster with traditional edge servers, with regards to electricity usage and monetary cost.
Our results quantitatively reveal when and for which applications mobile SoCs exhibit higher energy efficiency than traditional servers, as well as their ability to proportionally scale power consumption with fluctuating incoming loads.
These outcomes provide insightful implications and offer valuable direction for further refinement of the \soccluster to facilitate its deployment across wider edge scenarios.

\end{abstract}

	\section{Introduction}\label{sec:intro}

Energy efficiency has become a crucial factor in the design and operation of data centers~\cite{google-energy-proportional}.
According to reports~\cite{eu-dc-power}, EU data centers consumed 76.8 TWh of electricity in 2018, which accounted for 2.7\% of the total electricity usage within the EU.
This power consumption is estimated to increase to 98.52 TWh by 2030.
Such high energy usage also strains the cooling infrastructure and leads to significant costs for data center operators~\cite{asplos22-cooledge,osdi20-thunderbolt}.

Edge clouds, which provide computing resources in close proximity to users and devices, are becoming integral to daily life.
Their increasing deployment aims to reduce latency and enhance the performance of applications.
Nevertheless, energy issues are likely to intensify at the edge for several reasons.
First, power supplies to the edge are more limited and expensive, since edge servers are often near populated areas, unlike cloud data centers that may be strategically near abundant energy sources like hydro power.
Additionally, edge servers face spatial limitations and have a power density that is an order of magnitude higher than cloud servers.
This puts additional strain on cooling mechanisms~\cite{asplos22-cooledge}.
Moreover, edge workloads fluctuate more than cloud workloads due to the distinct features of edge applications~\cite{ens}.
One critical pathway towards sustainable data centers lies in enhancing the energy efficiency of individual servers.
Past efforts mainly focused on software optimizations~\cite{lee2021greendimm,liu2012renewable,ran2019deepee,manousakis2015coolprovision,zhang2021flex}, but hardware redesign may offer greater benefits~\cite{DBLP:conf/asplos/RanganathanSCDG21}.
However, implementing this can be challenging with established cloud infrastructure.
Fortunately, the emerging edge infrastructure presents a timely opportunity -- it is still in the final stage of development, and we are on the eve of its inauguration~\cite{shan2022towards,AWSexpandedge}.

In this study, we advocate for a new form of edge server, referred to as \soccluster, which comprises tens or hundreds of mobile SoCs, as an enhancement to existing edge infrastructure.
Mobile SoCs inherently possess higher energy efficiency than traditional datacenter servers, as they are designed for battery operated mobile devices with intermittent usage patterns~\cite{qcom-power-design,qcom-powereff}.
The \soccluster offers additional benefits, such as the ability to seamlessly run mobile operating systems and applications.
This benefit facilitates computation offloading, particularly in cloud gaming scenarios~\cite{nvidia-geforce,google-stadia,xcloud-games} for native mobile games.
Furthermore, mobile SoCs are inherently heterogenous.
With an optimized software stack, they can efficiently accommodate both general-purpose workloads (e.g., web services), and domain-specific tasks (e.g., deep learning~\cite{mnn,tflite} and multimedia processing~\cite{mediacodec}) by utilizing hardware accelerators such as GPUs and NPUs.

\begin{figure}[t]
	\centering					
	\includegraphics[width=0.45\textwidth]{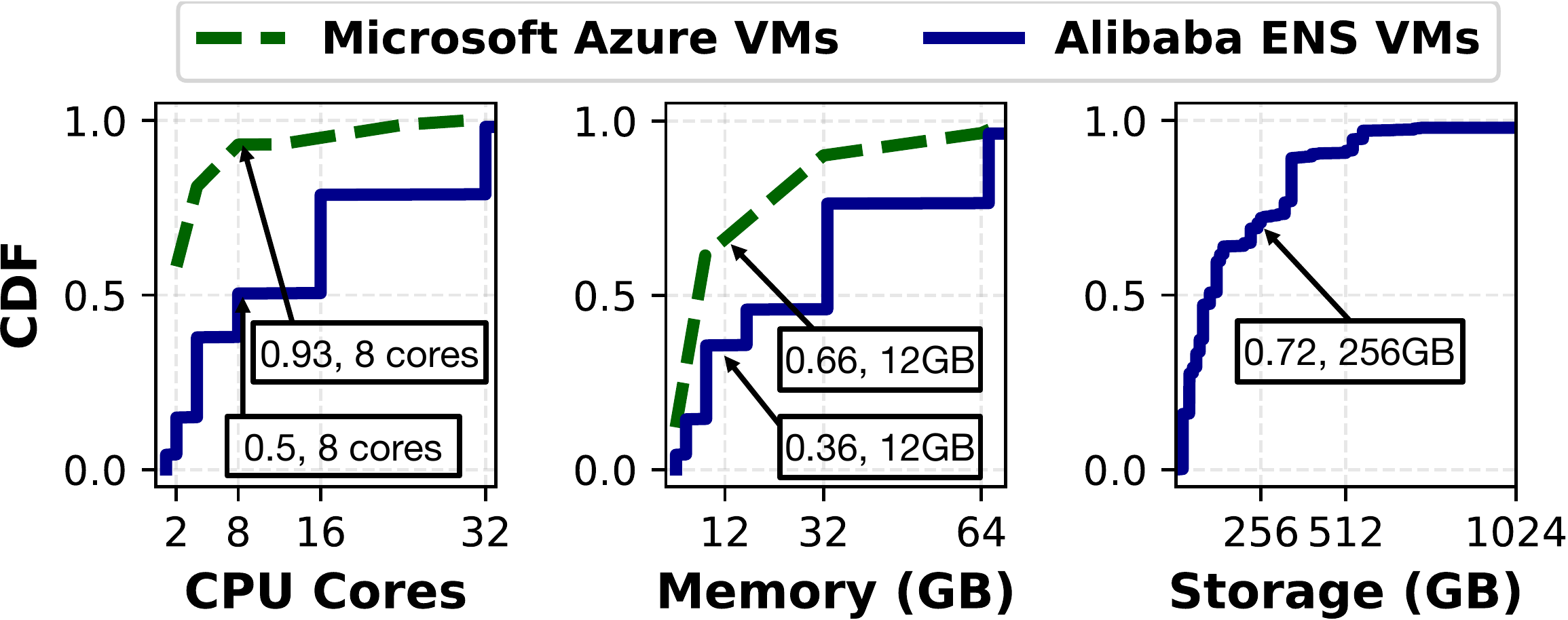}
	\caption{CDF of resource subscription of VMs in Microsoft Azure~\cite{cortez2017resource} and Alibaba ENS~\cite{ens}.
	\textmd{Approximately 66\% of Azure VMs and 36\% of Alibaba ENS VMs can be accommodated within a mobile SoC evaluated in this study (i.e., a Qualcomm Snapdragon 865 chip with 8 CPU cores, 12 GB memory and 256 GB storage).}
	}
	\vspace{-15pt}
	\label{fig:edgevm-specs}
\end{figure}

Breaking down a monolithic server into numerous smaller SoCs also caters to the fine-grained resource allocation required by cloud and edge applications.
Figure~\ref{fig:edgevm-specs} illustrates the resource subscription of 2.7 million VMs from Microsoft Azure~\cite{cortez2017resource} and 7,410 VMs from Alibaba ENS~\cite{ens}.
It reveals that most VMs on Azure/Alibaba datacenters require low resources, with up to 66\%/36\% of VMs having subscription configurations that fit within the hardware limits of a single mobile SoC (8 CPU cores, 12 GB memory, 256 GB storage).
The proportion is anticipated to grow in the future as the hardware improvement of mobile SoCs and the ongoing trend of software services shifting towards disaggregation~\cite{serverless-berkeley}.

\textbf{Hardware prototyping.}
We have developed a concrete \soccluster server, which integrates 60 Qualcomm Snapdragon 865 SoCs into a 2U rack, with detailed specifications described in $\S$\ref{sec:glance}.
Over the past two years, we have manufactured more than 10,000 such \socclusters, most of which have been deployed in edge clouds, primarily to serve cloud gaming workloads.
However, according to monitored traces ($\S$\ref{sec:soc-wild}), the utilization of the deployed \socclusters varies widely and is generally low.
Apparently the potential of these \socclusters has not been fully realized.
To fill the gap, the first and critical step is to demystify whether \socclusters can efficiently serve other edge applications.

\textbf{Measurement methodology.}
Therefore, we present a first-of-its-kind measurement study to quantitatively assess how our commercial-off-the-shelf \soccluster can efficiently support typical edge workloads.
This study focuses on two modern, computation-intensive workloads: video transcoding and deep learning (DL) serving.
Video transcoding stands as the predominant workload at the edge~\cite{ens}, with our experiments pinpointing two primary scenarios: live streaming transcoding and archive transcoding.
DL serving forms the essential component of numerous intelligent applications.
To provide a basis for comparison, we utilized a typical edge server with an Intel Xeon CPU and NVIDIA GPUs.
We expanded the testing to six mobile phones with high-end Qualcomm SoCs to broaden the findings.

\textbf{Benchmark suite.}
We developed a benchmark suite to test application performance on both the \soccluster and the traditional edge server.
The benchmark suite employs state-of-the-art libraries for each application.
For video transcoding, it uses FFmpeg~\cite{ffmpeg} to process six videos with disparate characteristics from vbench~\cite{vbench}.
For DL serving, it incorporates TFLite~\cite{tflite} (\soccluster), TVM~\cite{tvm} (Intel CPU), and TensorRT~\cite{tensorrt} (NVIDIA GPU).
The NN models used are ResNet-50, ResNet-152~\cite{he2015}, YOLOv5x~\cite{glenn_jocher_2022_6222936}, and BERT~\cite{bert-tfhub}.
We selected different software stacks for various hardware platforms as no single software optimally performs across heterogeneous processors.
The benchmark suite reports comprehensive metrics, including throughput, latency, and energy consumption under constraints like electricity and cost.

\textbf{Key findings} are summarized below.

\textit{\textbf{ (1) Energy efficiency.}}
The \soccluster demonstrates up to 6.5$\times$ higher throughput per unit of energy for serving DL inference workloads compared to the traditional edge server equipped with NVIDIA A40 GPUs.
Its energy efficiency is also comparable to high-end NVIDIA A100 GPUs.
However, for complex video transcoding workloads, SoC CPUs inside the \soccluster underperform to NVIDIA GPUs that are optimized for highly parallel tasks.
Additionally, attributing to the discrete SoC organization, the \soccluster can proportionally scale its energy consumption with dynamic loads, ensuring minimal degradation in energy efficiency.

\textit{\textbf{(2) Latency.}}
Due to a lack of proper software support for cross-SoC collaboration, the \soccluster faces challenges in handling delay-critical workloads.
While the inference latency of a single SoC for a medium-sized DNN model is low enough to meet the requirements of most edge applications (8.8 ms on quantized ResNet-50), the latency can be up to hundreds of milliseconds for larger models.
Therefore, there is an urgent need for a DL library to enable collaborative inference on multiple SoCs.

\textit{\textbf{(3) Monetary cost.}}
The \soccluster offers more than 2.23$\times$ greater throughput per monetary cost compared to the traditional edge server for live streaming transcoding.
On the other hand, NVIDIA GPUs significantly outperform the \soccluster in DL serving.
While this outcome might deter investment in new \socclusters for DL serving workloads, migrating lightweight or latency-insensitive DL tasks to the already deployed, underutilized \socclusters can still enhance energy efficiency.

\textit{\textbf{(4) SoC longitudinal study.}}
Through a longitudinal study, we found that mobile SoCs have demonstrated remarkable performance enhancements over the past six years, with a highest improvement of 8.5$\times$ on SoC DSPs.
Improvements in mobile co-processors position them as suitable candidates for handling more complex server-side workloads in the future.
Regarding the existing manufactured \socclusters, optimizing the current software stack is essential to fully utilize the capabilities of mobile co-processors.

\textbf{Contributions.} We made the following contributions.
\begin{itemize}
\item  We discussed the rationales and potential benefits of organizing mobile SoCs as edge servers, and presented the hardware prototyping and its commercial deployment by edge service providers.
\item  We designed and implemented a benchmark suite for two typical edge applications to evaluate their performance and power consumption on both the prototyped \soccluster server and a traditional edge server.
\item  We conducted a comprehensive measurement study based on the benchmark suite, and highlighted both the advantages and disadvantages of the \soccluster.
\item  We carried out a longitudinal study on various mobile SoCs to reveal their performance enhancements over time, and the potential to serve complex workloads with their co-processors.
\end{itemize}

	\section{Design and Prototyping}\label{sec:bkgnd}

\subsection{Motivations}

The current edge server architecture derives from the legacy of cloud computing that has been entrenched for decades.
Typically, an edge server consists of a many-core CPU along with a set of domain-specific accelerators (GPUs, TPUs, FPGAs, etc.).
Major edge resource providers worldwide, including Azure, AWS, and Alibaba, adhere to this design philosophy~\cite{aws-ec2-types,ali-ecs,azure-instance-types}.
However, the edge environment presents two distinctive characteristics that set it apart from the cloud:
(1) \textit{limited electricity and space availability}, resulting from the necessity to position edge servers close to populated areas, where the cost of electricity is high and physical spaces are confined;
(2) \textit{dynamic workloads} of user-oriented applications that challenge the energy efficiency of edge sites.
Although various software-level optimizations have been proposed~\cite{lee2021greendimm,manousakis2015coolprovision,atc20-serverless-trace}, we argue that it is time to reevaluate the architecture of edge servers.
In the meanwhile, the growing need to support mobile ecosystems in the cloud (e.g., native mobile games~\cite{gicloud} and virtual smartphone~\cite{cophone}) facilitates the adoption of mobile SoC-based servers.
In this work, we explore the feasibility and potential benefits of organizing multiple low-end SoCs as an edge server.

There exists some research on SoC servers.
Rajovic \textit{et al.} conducted an early analysis of utilizing mobile SoCs for high performance computing~\cite{rajovic2013supercomputing}.
Some studies focused on repurposing decommissioned mobile devices to reduce e-waste~\cite{shahrad2017towards,junkyardcomputing}.
Others explored the use of SoCs for specific applications, such as parallel computing~\cite{busching2012droidcluster}, key-value storage~\cite{andersen2009fawn}, web search~\cite{janapa2010web}, and video transcoding~\cite{liu2016greening}.
A similar vision to ours was presented in~\cite{sec22soc}, but with limited experiments and system design details. 
Those efforts typically include only theoretical analysis, or they conducted experiments on small-scale implementations involving 2--10 smartphones or Raspberry Pis, or evaluated simplified workloads~\cite{lang2010wimpy}.
Instead, our work aims to materialize the ``SoC-as-a-server'' concept in a more realistic and industry-relevant context.

\begin{figure}[t]
	\centering					
	\includegraphics[width=0.48\textwidth]{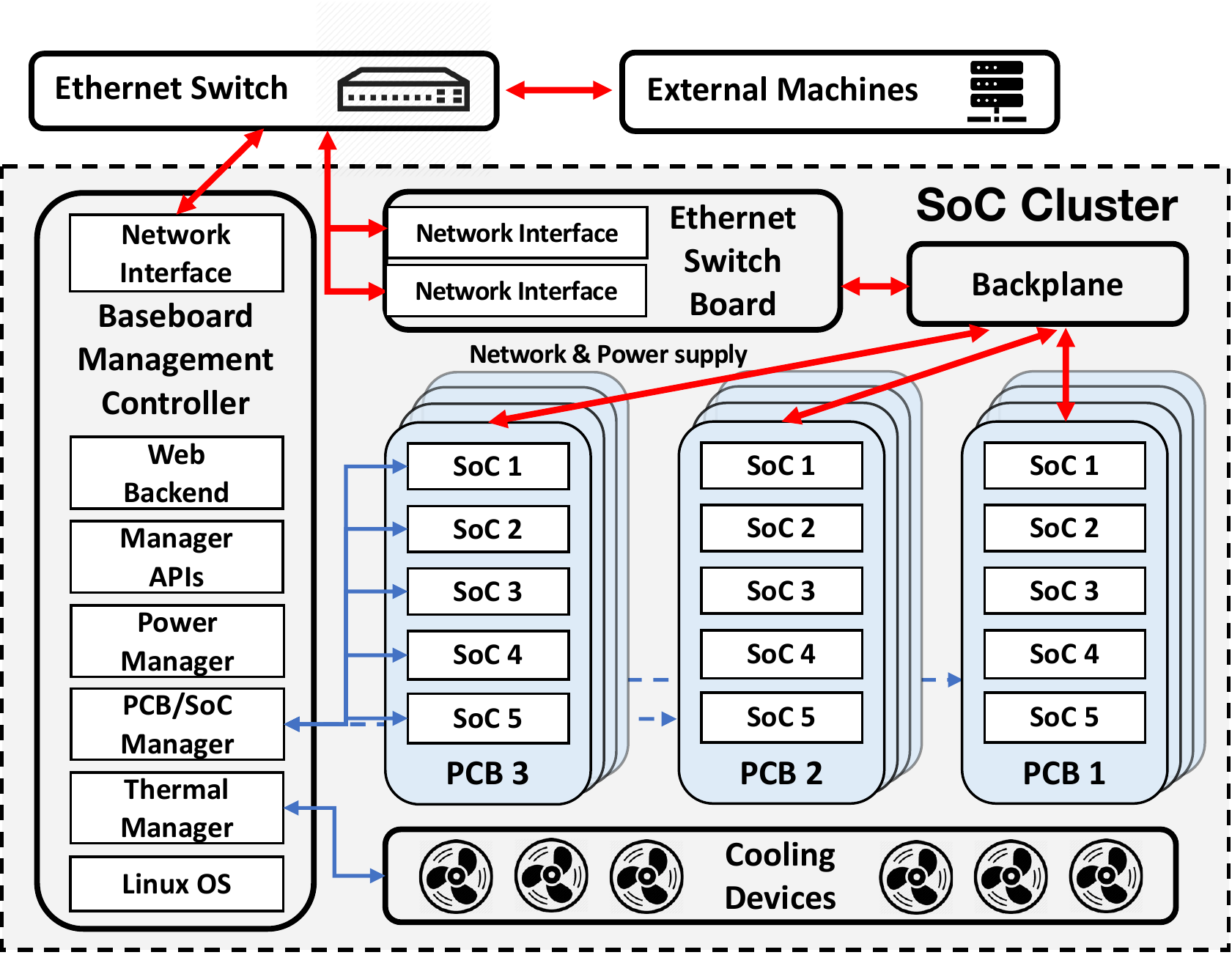}
	\caption{The architecture of \soccluster.}
	\vspace{-10pt}
	\label{fig:arch}
\end{figure}

\subsection{Implementing \soccluster}\label{sec:glance}

The design space for materializing the concept of an \soccluster into a server machine is expansive.
In this study, we design and implement a concrete \soccluster server that ensures maximal flexibility of each hardware component.
For simplicity, we refer to this concrete server as \soccluster.

\noindent \textbf{Overall architecture.}
Figure~\ref{fig:arch} illustrates the overall architecture of \soccluster.
The major component of the server is a pool of mobile SoCs, grouped into sets of five and integrated into individual printed circuit boards (PCBs).
These PCBs provide dedicated power supplies and network capabilities, and they act as network switches for interconnecting the SoCs.
Additionally, an Ethernet Switch Board (ESB) is incorporated into \soccluster to connect all SoCs to the external network.
The PCBs and ESB are interconnected through a backplane.
\soccluster also includes a Baseboard Management Controller (BMC) to monitor and manage server status, including power, temperature, and cooling devices.
In summary, the architecture of \soccluster is modularized, offering great flexibility in the design, manufacture, and upgrading.

\begin{figure}[t]
	\begin{minipage}[t]{0.48\textwidth}
	\begin{minipage}[t]{0.44\textwidth}
		\centering
		\includegraphics[width=0.79\textwidth]{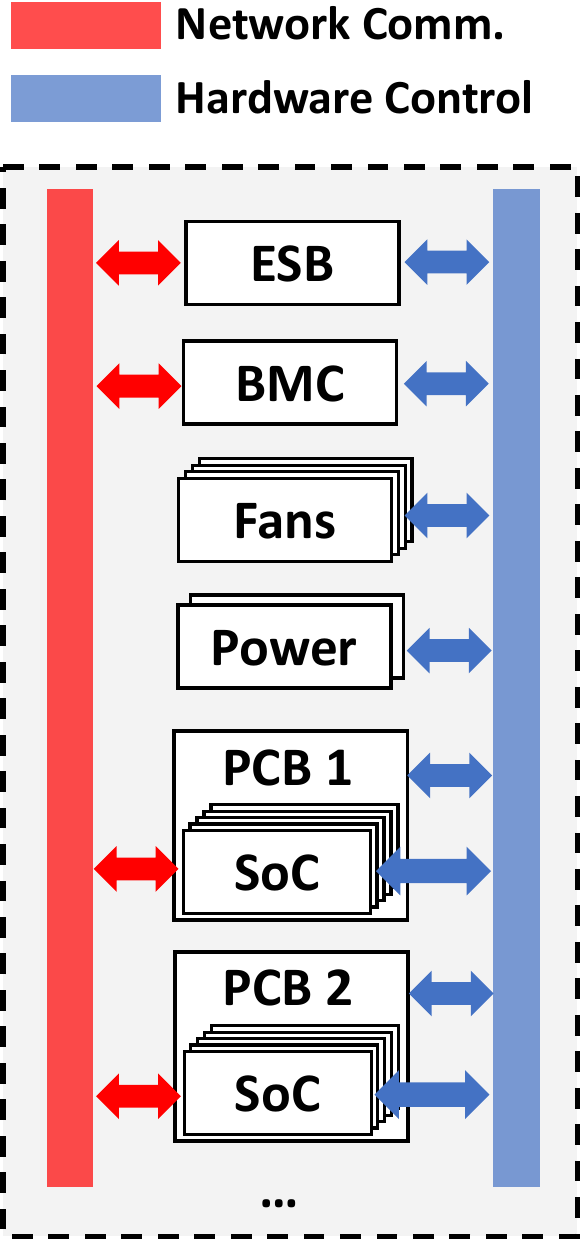}
		\caption{Hardware management in \soccluster.}
		\vspace{-10pt}
		\label{fig:hw-mgmt}
	\end{minipage}
	\hspace{3pt}
	\begin{minipage}[t]{0.5\textwidth}
		\centering
		\includegraphics[width=0.86\textwidth]{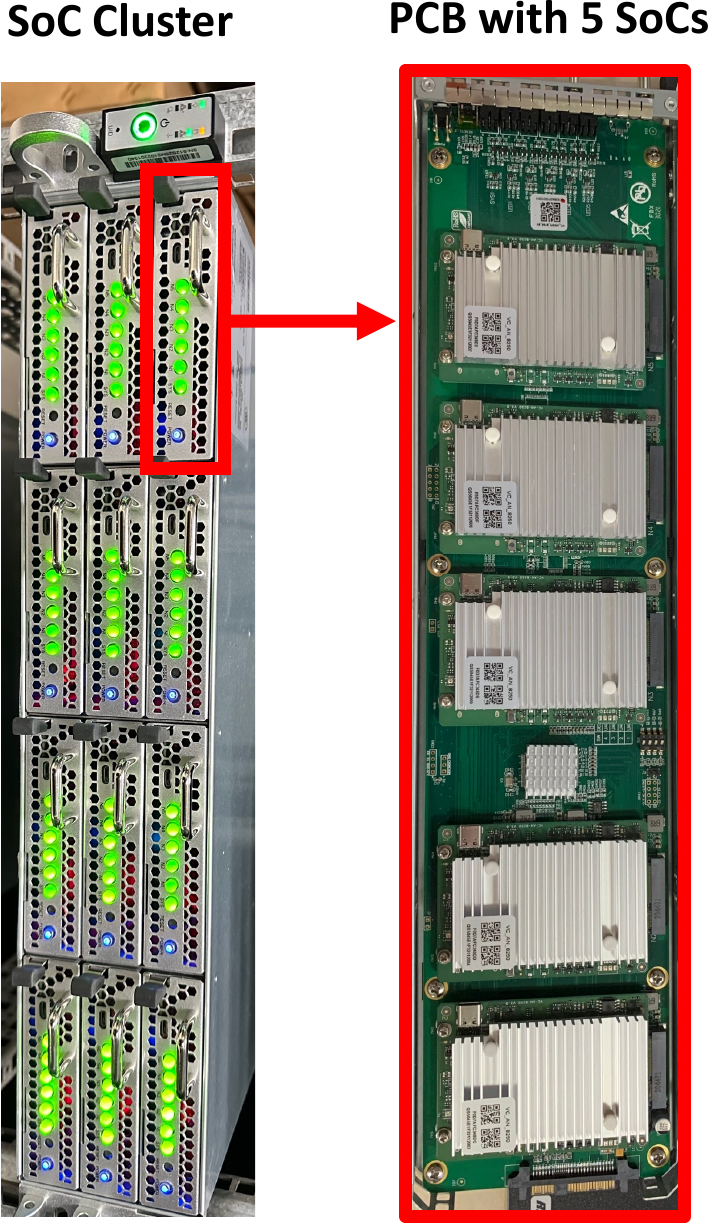}
		\caption{A manufactured \soccluster and a PCB.}
		\vspace{-10pt}
		\label{fig:real-hw}
	\end{minipage}
\end{minipage}
\end{figure}

\noindent \textbf{Hardware components and functionalities.}
Figure~\ref{fig:hw-mgmt} provides an overview of how the hardware components inside \soccluster are managed and connected.
Basically communication between two hardware components can be through hardware control messages or network-layer messages.
Here, we categorize the hardware components according to their functionalities.

\indent $\bullet$ \textit{Computing.}
\soccluster's computing units consist of 60 Qualcomm Snapdragon 865 SoCs~\cite{qcomm865-hardware}.
The number of integrated SoCs inside a \soccluster is mainly determined by the server's physical size and cooling capability.
Table~\ref{tab:spec} summarizes the detailed hardware and OS specifications of each SoC and the entire server.

\indent $\bullet$ \textit{Networking.}
The network functionality of \soccluster is primarily provided by two core hardware components.
The first is the ESB, which exposes the computing units to the external world through its dual SFP+ ports.
It supports up to 20 Gbps throughput.
The second component consists of 12 PCBs with network switching functionality. 
The Ethernet connection between the ESB and SoCs is relayed through the corresponding PCBs.
When a PCB is plugged into \soccluster, it establishes a physical connection with the ESB and builds an Ethernet connection with up to 1 Gbps throughput.

\indent $\bullet$ \textit{Management.}
\soccluster uses an independent hardware component, BMC, to monitor and control the computing units and all related server status, such as power supplies, temperature, and hardware failures.
The hardware control messages are transmitted through a mixture of protocols and technologies, including I2C, USB and UART.
The BMC also provides an Ethernet interface that enables external machines to access it.

\indent $\bullet$ \textit{Cooling and power supplies.}
To ensure proper cooling, \soccluster employs eight fans that circulate air through the mobile SoCs, the ESB, and the BMC, and then expel it from the fan module at the back of the server.
The server utilizes two power modules to provide redundant power supplies, with a maximum support of approximately 700 watts.

\begin{table}[t]
    \centering
    \footnotesize
    \includegraphics[width=0.45\textwidth]{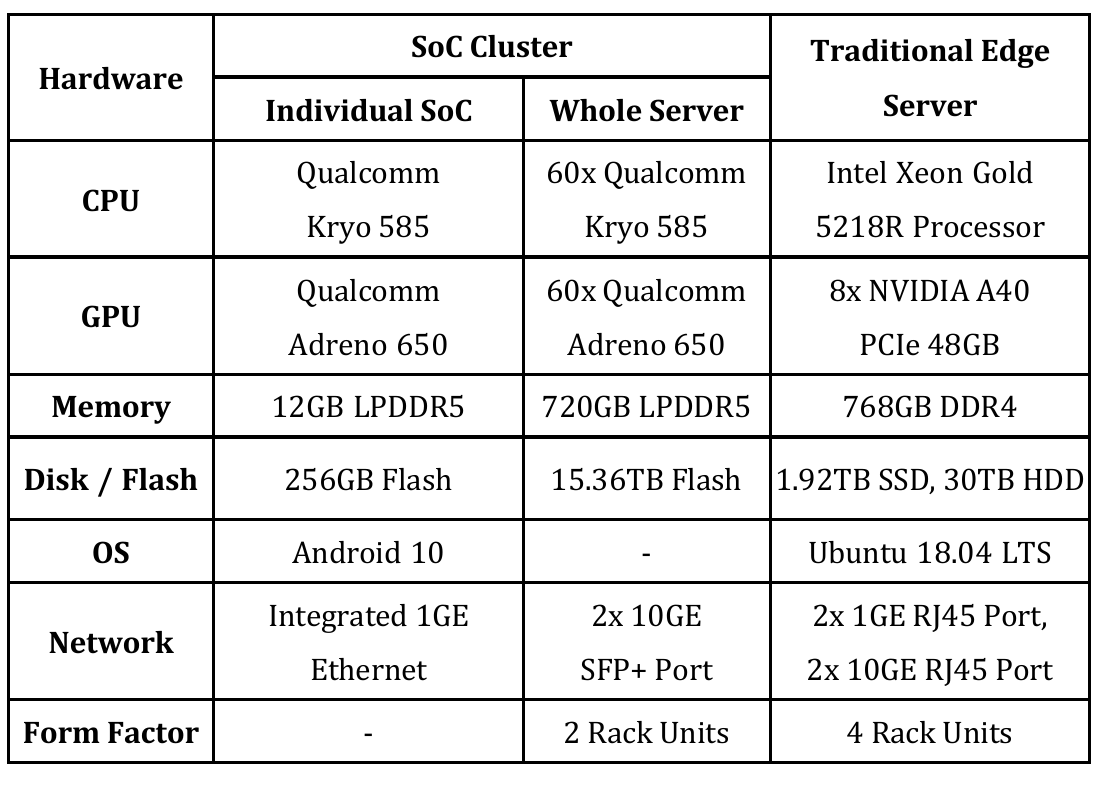}
    \caption{Two major hardware platforms used in this study. 
    }
    \label{tab:spec}
    \vspace{-5pt}
\end{table}

\subsection{Micro-experiments and Trace Analysis}\label{sec:soc-wild}

Figure~\ref{fig:real-hw} shows the manufactured \soccluster evaluated in this study, which has been on sale for over a year.
As of August 2023, more than 10,000 \socclusters have been shipped and deployed in the wild, primarily to edge service providers.
In collaboration with a leading worldwide edge service provider, we discovered that \textit{the deployed \socclusters are primarily used to serve a specific application: cloud gaming}.
Native mobile games, such as Genshin Impact~\cite{genshin}, usually release installation packages tailored for mobile platforms (e.g., for the arm64-v8a architecture).
While there exists mobile-in-cloud solutions designed for traditional servers with many-core CPUs and monolithic GPUs~\cite{anboxcloud,redroid}, these games can hardly run as smoothly as they do on mobile SoCs due to incompatibility with the CPU architecture and graphics stacks.
Recognized as the optimal hardware setup for native mobile games, the shipped \socclusters are estimated to serve millions of game sessions on a daily basis.

\begin{table}[t]
    \centering
    \scalebox{0.76}{
    \footnotesize
    \begin{tabular}{|l|cccc|cccc|}
    \hline
    \multicolumn{1}{|c|}{\multirow{2}{*}{\textbf{\begin{tabular}[c]{@{}c@{}}Micro\\Benchmarks\end{tabular}}}} & \multicolumn{4}{c|}{\textbf{Per-core Performance}}                                                                                & \multicolumn{4}{c|}{\textbf{Whole Server Performance}}                                                                            \\ \cline{2-9} 
    \multicolumn{1}{|c|}{}                                                                                            & \multicolumn{1}{c|}{\textbf{Ours}} & \multicolumn{1}{c|}{\textbf{Trad.}} & \multicolumn{1}{c|}{\textbf{G2}} & \textbf{G3} & \multicolumn{1}{c|}{\textbf{Ours}} & \multicolumn{1}{c|}{\textbf{Trad.}} & \multicolumn{1}{c|}{\textbf{G2}} & \textbf{G3} \\ \hline
    \textbf{CPU Score}                                                                                                & 911                                & 840                                 & 762                                  & 1,121           & 194,100                            & 15,450                              & 36,091                               & 51,379          \\ \hline
    \textbf{Integer Score}                                                                                            & 842                                & 800                                 & 735                                  & 1,039           & 184,500                            & 16,224                              & 36,653                               & 50,695          \\ \hline
    \textbf{Floating Score}                                                                                           & 948                                & 886                                 & 790                                  & 1,214           & 191,820                            & 15,793                              & 35,813                               & 49,885          \\ \hline
    \textbf{Text Compress}                                                                                     & 4.4                                & 4.1                                 & 4.2                                  & 4.9             & 906                                & 135                                 & 195                                  & 206             \\ \hline
    \textbf{SQLite Query}                                                                                         & 257                                & 249                                 & 208                                  & 279             & 59,958                             & 9,240                               & 12,200                               & 16,200          \\ \hline
    \textbf{PDF Render}                                                                                   & 52                                 & 41                                  & 37                                   & 66              & 12,552                             & 710                                 & 2,140                                & 3,960           \\ \hline
    \end{tabular}
    }
    \caption{A list of micro-benchmark (from Geekbench 5~\cite{geekbench}) results on \soccluster and typical edge servers.
    \textmd{
    ``Ours'': \soccluster;
    ``Trad.'': the traditional edge server;
    ``G2/3'': AWS Graviton 2/3 cloud instances (m6g.metal/m7g.metal; both with 64 cores, 256 GB RAM);
    }
    }
    \vspace{-15pt}
    \label{tab:microbenchmark}
\end{table}

\noindent \textbf{Micro-benchmarks on CPU.}
To initiate an understanding of \soccluster's performance, we employed a cross-platform benchmark, Geekbench 5~\cite{geekbench}, to run a series of CPU-only micro-benchmarks on four different servers, 
including \soccluster and the traditional edge server as shown in Table~\ref{tab:spec}, and AWS Graviton 2/3 servers with ARM CPUs designed for cloud environments.
The results in Table~\ref{tab:microbenchmark} provide the following observations.
First, the per-core performance of \soccluster aligns closely with that of the Intel Xeon CPU, outperforming the AWS Graviton 2 processor.
Second, from a whole server's perspective, the large number of integrated SoCs grants \soccluster superior performance compared to other CPU servers.
For example, it exhibits 3.8$\times$ higher CPU core score and 3.2$\times$ faster PDF rendering speed relative to the latest AWS Graviton 3 cloud instance.
These results demonstrate the significant potential of SoC Cluster's applicability for general edge workloads.

\noindent \textbf{Network performance.}
We used ping and iPerf3 to assess network round-trip time (RTT) and TCP/UDP bandwidth between two individual SoCs.
The results show that the network is stable, with an RTT of approximately 0.44 ms and TCP and UDP bandwidths of nearly 903 Mbps and 895 Mbps, respectively.
However, these results do not truly reflect the performance of the networking system, as they were not measured on real edge applications and did not involve multiple SoCs simultaneously.
In subsequent sections, we further investigate the network performance (live streaming transcoding in $\S$\ref{sec:video-network} and cross-SoC DL serving in $\S$\ref{sec:dl-collaborative}).

\begin{figure}[h]
	\centering					
	\includegraphics[width=0.46\textwidth]{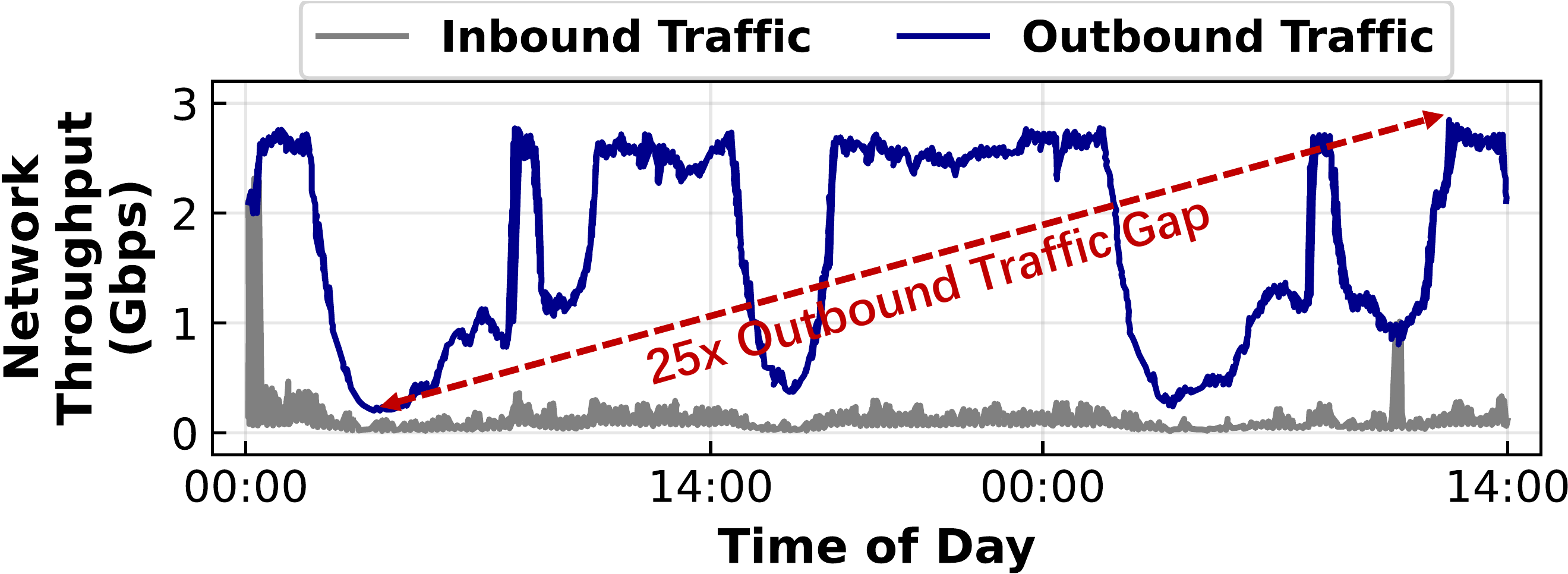}
	\caption{The network throughput of an in-the-wild \soccluster that serves cloud gaming workloads over 38 hours. 
	\textmd{The server is randomly picked from an edge site of one edge service provider. 
	Full network capacity: 20 Gbps.}
	}
	\label{fig:usage-var}
\end{figure}

\noindent \textbf{Trace analysis.}
Although \soccluster exhibits superior performance at the whole server level, our analysis of runtime traces collected from deployed \socclusters reveals that \textit{these servers experience workloads with relatively low utilization and high dynamics.}
In general, the resource usage of all deployed \socclusters remains below 20\%.
Figure~\ref{fig:usage-var} displays the temporal network traffic of a specific \soccluster randomly selected from real-world edge sites.
As observed, the disparity between its highest and lowest outbound traffic reaches up to 25$\times$.
This observation aligns with a recent empirical study on large-scale edge clouds~\cite{ens}, which discovered that edge applications are mainly user-centric, therefore highly dependent on user activities.
Motivated by this insight, this study seeks to explore how efficiently \soccluster can serve other typical edge workloads beyond cloud gaming.

	\section{Methodology and Benchmark}\label{sec:design}

In this work, we perform an application-driven measurement study to demystify the performance of \soccluster and traditional edge servers.
This section elaborates on the methodology used to set up the experiments, with best possible effort to ensure a fair comparison.

\noindent \textbf{Applications.}
We select two modern, computation-intensive applications:
(1) \textit{Video transcoding}~\cite{video-tr}, which involves converting the format (FPS, resolution, etc.) of a given video stream, is widely adopted in online conferences and live streaming.
It is reported to be a dominant workload at the edge~\cite{li2023demystifying,ens}.
We identify two specific scenarios for this workload: live streaming transcoding and archive transcoding, with their characteristics illustrated in $\S$\ref{sec:video-trans}.
(2) \textit{DL serving} is a key component in intelligent applications like AR, VR, and autonomous driving~\cite{tf-serving,torch-serving,nv-serving}.
A DL serving system receives a stream of input data and executes it using a designated DL model.
Substantial academic effort has been directed at optimizing DL serving performance~\cite{atc17-clipper,osdi20-serving,atc19-mark,fb-inference}.
In summary, the two applications cover the interests of both industry and academia, representing the major workloads at the edge.
They are both resource-intensive, indicating that if \soccluster performs well on them, it has the potential to efficiently serve other applications as well.

\noindent \textbf{Metrics}
are divided into two categories: application performance metrics and comparison metrics.
The former are mainly about throughput (the number of processed samples or videos per second), latency (model inference time), and power consumption.
To ensure a relatively fair comparison between SoC Cluster and the traditional edge server, we further judiciously select two conditional constraints:

\indent $\bullet$ \textit{Energy.}
As previously mentioned, energy consumption is a critical constraint for edge sites.
Therefore, we use throughput per energy unit (e.g., Joule) to assess application energy efficiency.
Besides energy efficiency under a specific (often full) load, we also consider energy proportionality under various load levels, acknowledging that edge servers experience high load variations.
An ideal edge server should scale its power consumption proportionally with the load to minimize wasted energy~\cite{google-energy-proportional}.

\indent $\bullet$ \textit{Monetary cost.}
We conduct a total cost of ownership (TCO) analysis to provide insight into the relationship between application performance and monthly TCO.
This analysis is designed to assist edge operators in making well-informed purchasing and scheduling decisions.

\noindent \textbf{Hardware.}
The \soccluster used in measurements was introduced in $\S$\ref{sec:glance}.
For comparison, we use a traditional server equipped with an Intel Xeon Gold CPU (4.0 GHz and 40 physical cores), 8 NVIDIA A40 GPUs, and 768 GB DRAM, running Ubuntu 18.04 LTS;
its specifications are summarized in Table~\ref{tab:spec}.
We confirmed that this type of server is widely used at the edge sites where \socclusters are deployed.
In our DL serving experiments, we opt for a higher-end NVIDIA A100 GPU from the Google Cloud Platform for a more comprehensive comparison.
The decision was made since NVIDIA A40 is not the highest-end server-level GPU released in 2020, the same year as the Qualcomm Snapdragon 865 SoC.
We exclude the NVIDIA A100 GPU in video transcoding experiments due to its lack of support for NVENC video encoding as of May 2024~\cite{nv-support-matrix}.

\noindent \textbf{Benchmark suite.}
One challenge of this study is selecting the appropriate software stack.
Given the inherently heterogeneous hardware architectures, we found that there is rarely software compatible with each processor type (i.e., SoC CPU/GPU/DSP, Intel CPU, and NVIDIA GPU).
Even if such software exists, its performance could be far from state-of-the-art standards.
To obtain meaningful results, we consulted with our industry partners and conducted testing with commonly used software.
We then selected best-performing option for each application and hardware configuration.

\indent $\bullet$ \textit{Video transcoding.}
We use FFmpeg (v4.4)~\cite{ffmpeg} with libx264~\cite{x264} and NVDEC/NVENC~\cite{nvcodec} support.
We cross-compiled FFmpeg to \soccluster with ARMv8 NEON acceleration.
We employ a popular open-source Android library, LiTr~\cite{litr}, for hardware-accelerated transcoding on \soccluster, due to FFmpeg's limited support for Qualcomm SoCs.
The benchmark suite is built atop vbench~\cite{vbench}, a benchmark tool widely used for cloud video transcoding.

\indent $\bullet$ \textit{DL serving}.
We employ TFLite~\cite{tflite} for \soccluster, TVM~\cite{tvm} for the Intel CPU, and TensorRT~\cite{tensorrt} for the NVIDIA GPU.
We select one medium-sized DNN (ResNet-50~\cite{he2015}) and three large DNNs (ResNet-152~\cite{he2015}, YOLOv5x~\cite{glenn_jocher_2022_6222936}, and BERT base~\cite{bert-tfhub}), all of which are representative DL serving workloads.

\noindent \textbf{Setups.}
For experiments on the Intel CPU, we partition the 80 cores (80 hardware threads) into 10 separate 8-core Docker containers.
We select 8 cores for two reasons.
First, previous edge workloads~\cite{ens} have shown that 8 is the median number of vCPU cores for edge IaaS VMs, which is also adequate for most edge services.
Second, the SoC's CPU also contains 8 cores, making the comparison more direct.
We leverage the \texttt{turbostat} command to read CPU/RAM power and the \texttt{nvidia-smi} command to access GPU power on the traditional edge server.
On \soccluster, we utilize BMC's API (implemented atop the I2C protocol) to measure power consumption of the whole server.
Our report on workload power consumption excludes idle power.
By default, our experiments are conducted with hardware fully loaded by batching DL serving or starting multiple transcoding processes.
We vary the load levels only when testing the energy proportionality.

Furthermore, to reduce power fluctuations during experiments:
(1) in live streaming transcoding, we start the maximum number of streams supported by each hardware, ensuring that no stream's performance (FPS) fell below that of the origin video stream;
(2) in archive transcoding, we repeat the same transcoding process on a video ten times;
(3) in DL serving, we set the DL inference iteration to 1,000 for each test.
	\section{Video Transcoding Results}\label{sec:video-trans}
\begin{table*}[t]
	\centering
	\scalebox{0.9}{
	\footnotesize
	\begin{tabular}{|l|rrrrr|rrr|}
		\hline
		\multicolumn{1}{|c|}{\multirow{2}{*}{\textbf{Video}}} & \multicolumn{5}{c|}{\textbf{Video Metadata}} & \multicolumn{3}{c|}{\textbf{Network Bound Analysis of Live Streaming Transcoding}} \\ \cline{2-9} 
		\multicolumn{1}{|c|}{} & \multicolumn{1}{c|}{\textbf{Resolution}} & \multicolumn{1}{c|}{\textbf{FPS}} & \multicolumn{1}{c|}{\textbf{\begin{tabular}[c]{@{}c@{}}Source\\ Entropy\end{tabular}}} & \multicolumn{1}{c|}{\textbf{\begin{tabular}[c]{@{}c@{}}Source\\ Bitrate\end{tabular}}} & \multicolumn{1}{c|}{\textbf{\begin{tabular}[c]{@{}c@{}}Target\\ Bitrate\end{tabular}}} & \multicolumn{1}{c|}{\textbf{\begin{tabular}[c]{@{}c@{}}Max. Stream Num\\ (per SoC)\end{tabular}}} & \multicolumn{1}{c|}{\textbf{\begin{tabular}[c]{@{}c@{}}Max. Network Usage\\ (per PCB, 1 Gbps)\end{tabular}}} & \multicolumn{1}{c|}{\textbf{\begin{tabular}[c]{@{}c@{}}Max. Network Usage\\ (whole server, 20 Gbps)\end{tabular}}} \\ \hline
		V1: holi & \multicolumn{1}{r|}{854x480} & \multicolumn{1}{r|}{30} & \multicolumn{1}{r|}{7.0} & \multicolumn{1}{r|}{2.8 Mbps} & 819.8 Kbps            & \multicolumn{1}{r|}{13 (CPU) / 16 (HW)} & \multicolumn{1}{r|}{534 Mbps (53.4\%)   } & 6,407 Mbps (32.0\%)  \\ \hline
		V2: desktop & \multicolumn{1}{r|}{1280x720} & \multicolumn{1}{r|}{30} & \multicolumn{1}{r|}{0.2} & \multicolumn{1}{r|}{181 Kbps} & 90.5 Kbps         & \multicolumn{1}{r|}{15 (CPU) / 16 (HW)} & \multicolumn{1}{r|}{43 Mbps (4.3\%)     } & 505 Mbps (2.5\%)     \\ \hline
		V3: game3 & \multicolumn{1}{r|}{1280x720} & \multicolumn{1}{r|}{59} & \multicolumn{1}{r|}{6.1} & \multicolumn{1}{r|}{5.6 Mbps} & 2.7 Mbps            & \multicolumn{1}{r|}{4 (CPU) / 12 (HW) } & \multicolumn{1}{r|}{673 Mbps (67.3\%)   } & 8,072 Mbps (40.3\%)  \\ \hline
		V4: presentation & \multicolumn{1}{r|}{1920x1080} & \multicolumn{1}{r|}{25} & \multicolumn{1}{r|}{0.2} & \multicolumn{1}{r|}{430 Kbps} & 215 Kbps    & \multicolumn{1}{r|}{9 (CPU) / 16 (HW) } & \multicolumn{1}{r|}{81 Mbps (8.1\%)     } & 968 Mbps (4.8\%)     \\ \hline
		V5: hall & \multicolumn{1}{r|}{1920x1080} & \multicolumn{1}{r|}{29} & \multicolumn{1}{r|}{7.7} & \multicolumn{1}{r|}{16 Mbps} & 4.1 Mbps             & \multicolumn{1}{r|}{3 (CPU) / 7 (HW)  } & \multicolumn{1}{r|}{1,008 Mbps (100.8\%)} & 12,010 Mbps (60.5\%) \\ \hline
		V6: chicken & \multicolumn{1}{r|}{3840x2160} & \multicolumn{1}{r|}{30} & \multicolumn{1}{r|}{5.9} & \multicolumn{1}{r|}{49 Mbps} & 16.6 Mbps         & \multicolumn{1}{r|}{1 (CPU) / 2 (HW)  } & \multicolumn{1}{r|}{985 Mbps (98.5\%)   } & 11,821 Mbps (59.1\%) \\ \hline
	\end{tabular}
	}
\caption{The metadata of videos in the transcoding experiments and the network bound analysis of live streaming transcoding on \soccluster.
\textmd{The videos are picked from vbench~\cite{vbench} to ensure diverse coverage of resolution, FPS, and entropy.
Entropy is calculated by bits per pixel per second and thus relating to the scene complexity.
The network usages include both inbound and outbound traffic.
CPU and HW represent transcoding on SoC CPU and SoC hardware codec.
}
}
\vspace{-10pt}
\label{tab:transcoding-task}
\end{table*}

In this section, we compare video transcoding performance in two scenarios with distinct characteristics: live streaming transcoding and archive transcoding.
The former usually takes video streams with a constant frame rate, widely used in live streaming and online conferences.
The latter processes video clips from file storage, typically serving as a preliminary stage before distribution to content providers~\cite{youtube-upload,vbench}.
Following the prior video benchmark~\cite{vbench}, we maintain a constant target bitrate for live streaming transcoding and consistent video quality for archive transcoding across hardware.
The metadata for the videos is shown in Table~\ref{tab:transcoding-task}.

\subsection{Energy Efficiency}\label{sec:video-energy}
\begin{figure}
	\centering					
	\begin{minipage}[t]{0.48\textwidth}
		\includegraphics[width=1\textwidth]{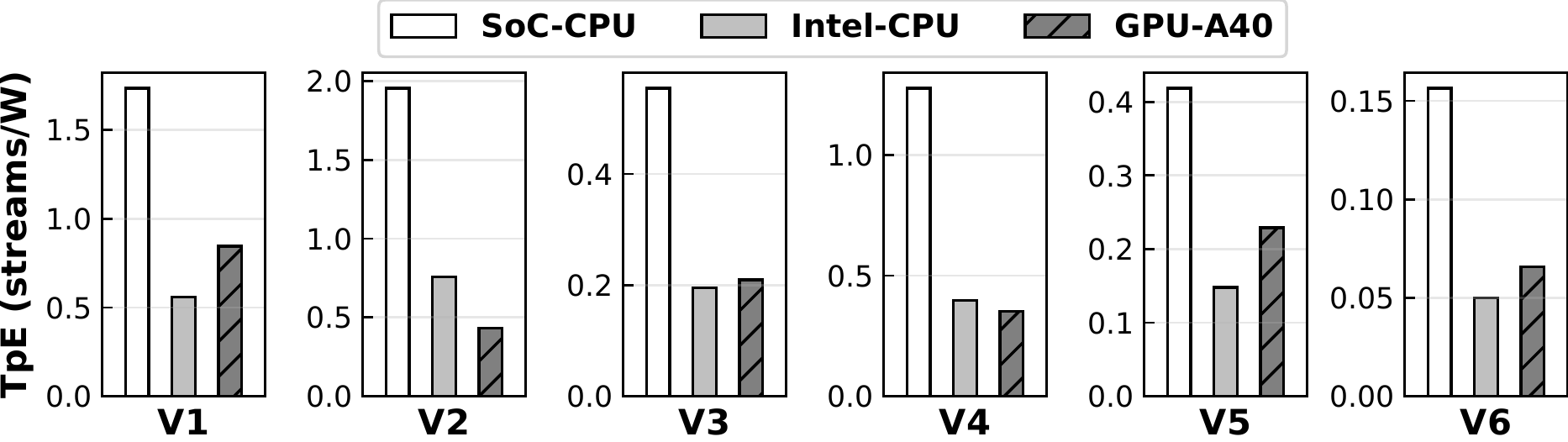}
		\subcaption{Live streaming transcoding}
		\label{fig:video-on-tpe}
	\end{minipage}
	
	\begin{minipage}[t]{0.48\textwidth}
		\includegraphics[width=1\textwidth]{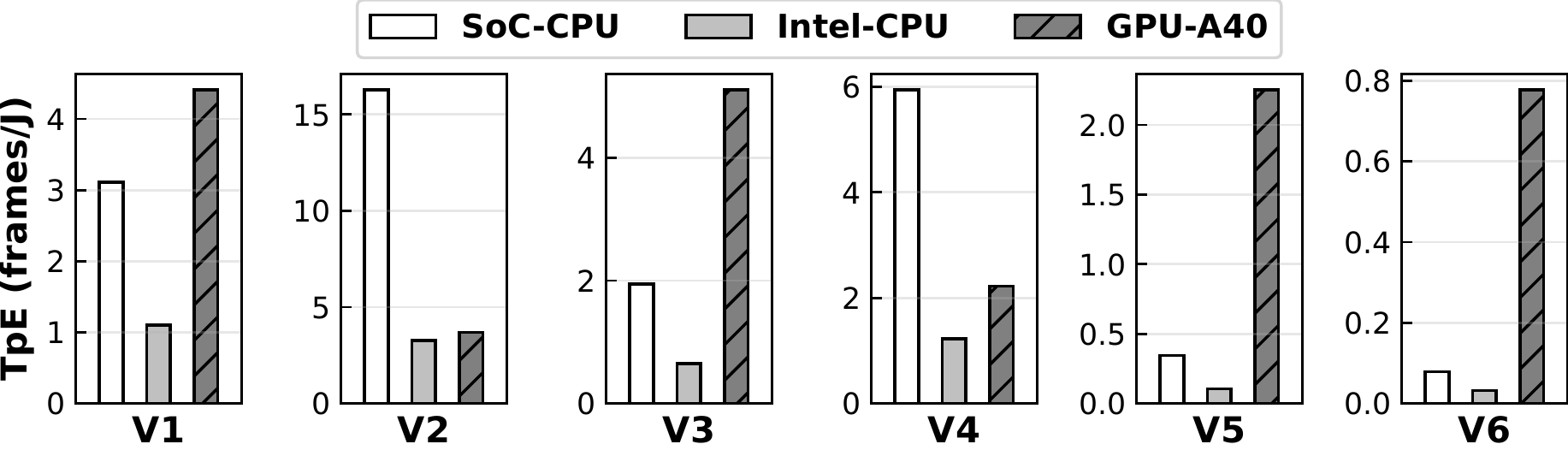}
		\subcaption{Archive transcoding}
		\label{fig:video-off-tpe}
	\end{minipage}
	\caption{Transcoding energy efficiency.}
	\vspace{-10pt}
\end{figure}

For live streaming transcoding, energy efficiency is measured as the number of video streams supported per watt (streams/W).
For archive video transcoding, energy efficiency is measured as how many frames can be processed per Joule (frames/J).
We unify the above metrics as throughput per energy unit (TpE).
Figure~\ref{fig:video-on-tpe} and \ref{fig:video-off-tpe} present the results for live streaming transcoding and archive transcoding, respectively.
Our key observation is that \textit{\soccluster shows significant energy savings across all live streaming and partial archive transcoding tasks over the Intel CPU and the NVIDIA GPU.}

In live streaming transcoding, \soccluster's SoC CPUs are 2.58$\times$--3.21$\times$ more energy-efficient than the Intel CPU, and 1.83$\times$--4.53$\times$ more energy-efficient than the NVIDIA A40 GPU across different videos.
In archive transcoding, SoC CPUs consistently outperform the Intel CPU in energy efficiency, although their advantage over the NVIDIA GPU varies by video.
Specifically, the NVIDIA GPU performs worse on videos V2 and V4.
We find the common feature of V2 and V4 is that they have low entropy due to minimal motion or rare scene transitions.
In such cases, we observe that the NVIDIA GPU stays in a high-power mode with high clock frequencies, while SoC CPUs accomplish these tasks using minimal CPU resources.
The similar observation can be proven in live streaming transcoding tasks -- \soccluster demonstrates superior energy efficiency on videos with lower complexity.

\begin{figure}[t]
	\centering
		\includegraphics[width=0.45\textwidth]{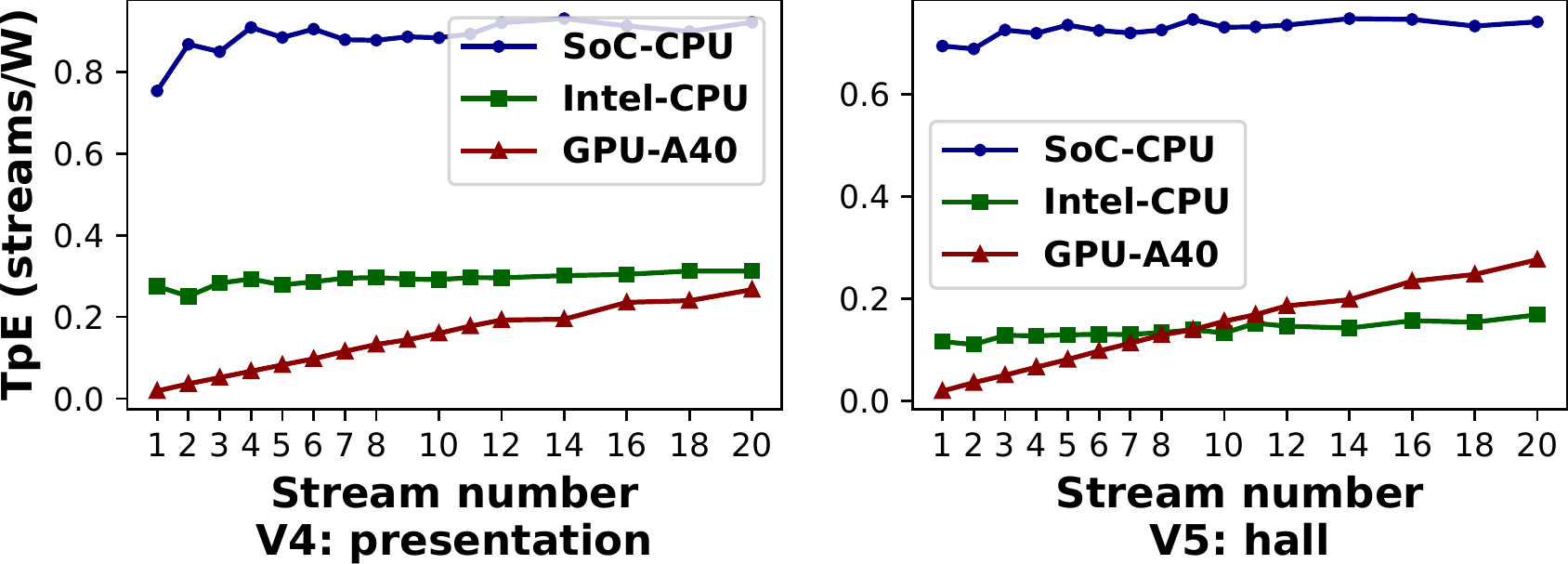}
		\vspace{-5pt}
		\caption{Energy efficiency of live streaming transcoding with different numbers of live video streams being processed simultaneously.}
		\vspace{-15pt}
		\label{fig:video-power-efficiency}
\end{figure}

We further explore how energy efficiency scales with dynamic workloads at the edge.
To conduct this analysis, we manually adjust the number of simultaneous transcoded video streams from 1 to 20 on all hardware.
We use two 1080p videos with diverse scene complexity and present the relationship between throughput per watt and the number of processed video streams.
Figure~\ref{fig:video-power-efficiency} shows a consistent trend for both videos:
both SoC CPUs and the Intel CPU display nearly constant energy efficiency as the number of video streams rises, implying a linear increase in power consumption with increasing workloads.
As for the NVIDIA GPU, it can only process 0.018 live video streams per watt when transcoding a single video (V4), falling 14.9$\times$ behind the Intel CPU and 40.8$\times$ behind SoC CPUs.
As the number of video streams increases, energy efficiency of the NVIDIA GPU gradually increases but is still lower than that of SoC CPUs.
The fine-grained control over SoC CPU cores or SoCs offers better energy scalability for serving dynamic video transcoding workloads on \soccluster.

\begin{figure}[t]
	\centering
	\begin{minipage}{0.45\textwidth}
		\includegraphics[width=1\textwidth]{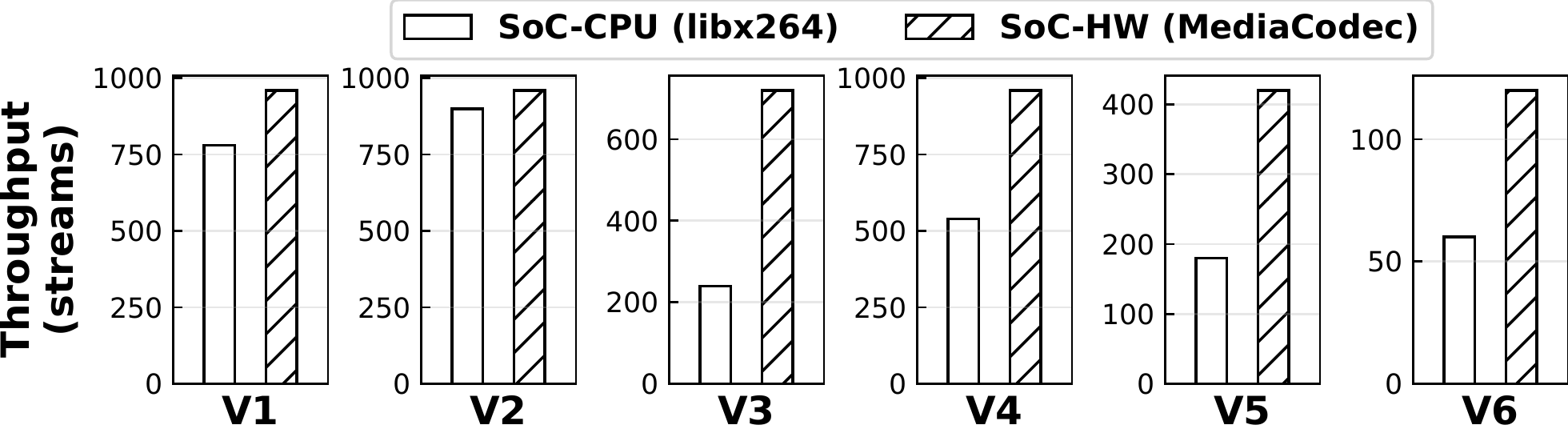}
		\subcaption{Transcoding throughput of the whole \soccluster.}
		\label{fig:video-mediacodec-tps}
	\end{minipage}
	
	\vspace{5pt}
	\begin{minipage}{0.45\textwidth}
		\includegraphics[width=1\textwidth]{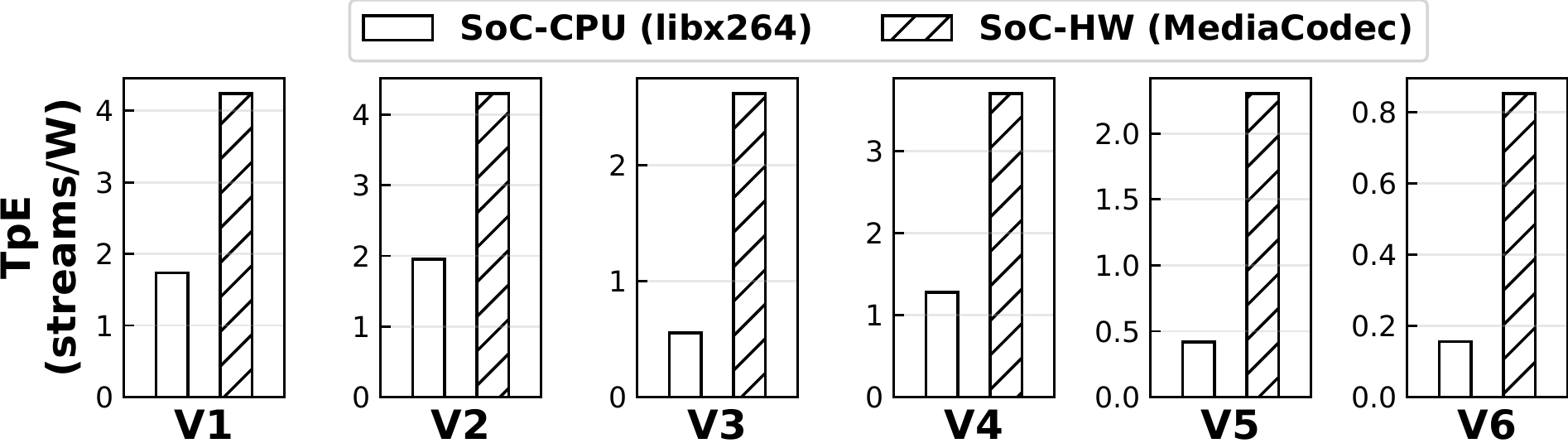}
		\subcaption{Energy efficiency.}
		\label{fig:video-mediacodec-tpe}
	\end{minipage}

	\vspace{-5pt}
	\caption{Live streaming transcoding performance of SoC CPU and hardware codec in \soccluster.}
	\vspace{-18pt}
\end{figure}

\subsection{Hardware-accelerated Video Transcoding on SoCs}\label{sec:video-bitrate}
Mobile SoCs are equipped with hardware codecs to handle most of the video encoding/decoding tasks on smartphones.
Our tests focus on live streaming transcoding because Android MediaCodec's APIs~\cite{mediacodec} lack the controls for video quality, which is essential for a fair comparison in archive transcoding.
We used LiTr~\cite{litr} for hardware-accelerated video transcoding instead, given that FFmpeg only supports hardware decoding but not encoding on the Android platform~\cite{ffmpeg-hwaccel}.
Below, we compare the performance and transcoding behavior of the hardware codec against SoC CPU, focusing on key metrics such as server-side transcoding throughput, energy efficiency, output bitrate, and video quality.

Our key observation is that \textit{the hardware codec of the mobile SoC exhibits a significant improvement over its CPU.}
As shown in Figure~\ref{fig:video-mediacodec-tps}, employing the hardware codec increases the maximum number of supported live video streams by 1.07$\times$--3$\times$.
The increase in energy efficiency is even more impressive, as illustrated in Figure~\ref{fig:video-mediacodec-tpe}.
For videos with low-complexity scenes (V1, V2, and V4), the hardware transcoders in \soccluster can support a geometric mean of 2.5$\times$ more streams per watt compared to SoC CPUs.
When transcoding high-entropy and high-resolution videos (i.e., V3, V5, and V6), SoC CPUs consume more power for video encoding, while offloading transcoding workloads to hardware codecs results in a significant boost in energy efficiency, with improvements ranging from 4.7$\times$ to 5.5$\times$.

\begin{figure}[t]
	\centering
	\begin{minipage}{0.45\textwidth}
		\includegraphics[width=1\textwidth]{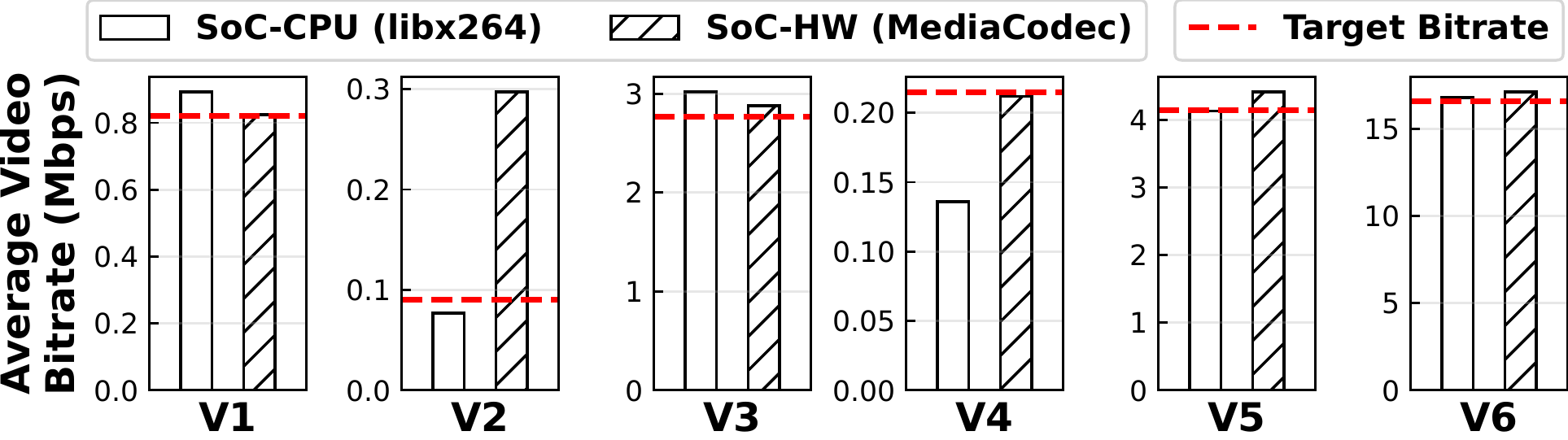}
		\vspace{-15pt}
		\caption{Target/output video bitrate in live streaming transcoding.
		}
		\label{fig:video-mediacodec-bitrate}
	\end{minipage}

	\begin{minipage}{0.45\textwidth}
		\includegraphics[width=1\textwidth]{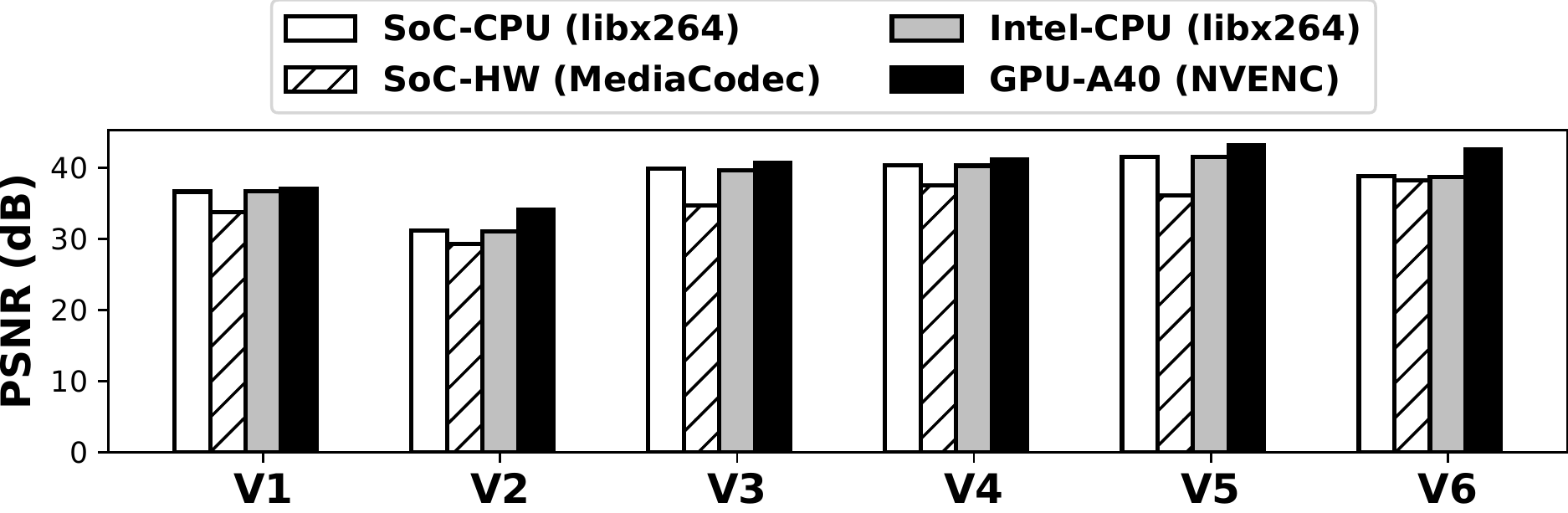}
		\vspace{-10pt}
		\caption{Live streaming transcoding quality of different encoders with the same bitrate constraint.}
		\vspace{-10pt}
		\label{fig:video-enc-quality}
	\end{minipage}
\end{figure}

To further clarify the ability of hardware codecs in \soccluster to handle live streaming transcoding workloads, we evaluate the output video bitrate, which is one of the most critical metrics affecting user experience.
As depicted in Figure~\ref{fig:video-mediacodec-bitrate}, we use red dash lines to indicate the target bitrate for each transcoding task.
The primary finding reveals that, \textit{in most cases, the hardware codec can meet the bitrate constraint, but it struggles to meet a relatively low bitrate cap.}
For example, setting a target bitrate of 90.5 Kbps for V2 will make the encoder create a higher bitrate output (even higher than the origin video stream).
Such unexpected behavior runs counter to the typical objective of archive transcoding services to compress a video stream.
The same behaviors were confirmed by our supplementary experiments on other videos and ultra-low bitrate settings.
This may suggest a potential design trade-off made by the mobile SoC vendor with considerations for energy efficiency and chip size.

\begin{figure*}[t]
	\centering
	\begin{minipage}[]{0.95\textwidth}
		\includegraphics[width=1\textwidth]{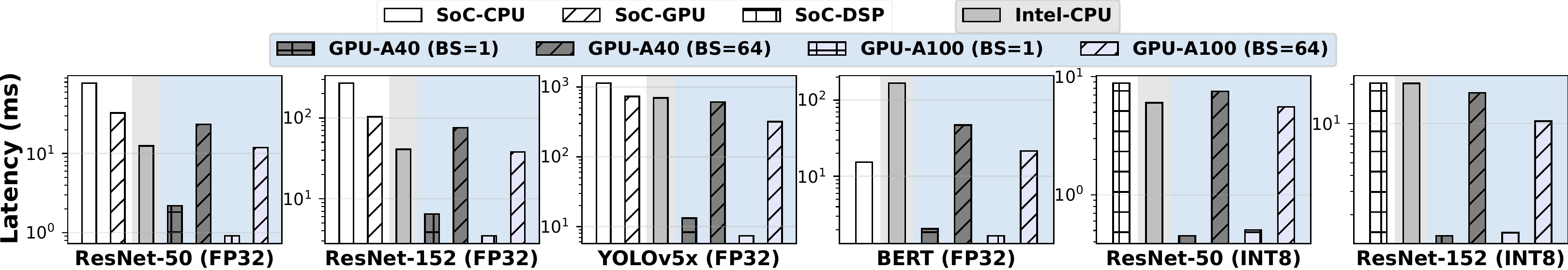}
		\subcaption{Deep learning serving latency.
		}
		\label{fig:dl-latency}
	\end{minipage}
	
	\begin{minipage}[]{0.95\textwidth}
		\includegraphics[width=1\textwidth]{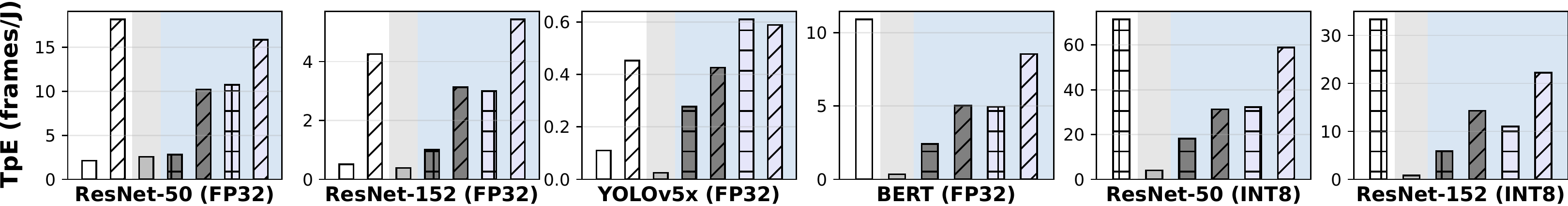}
		\subcaption{Deep learning serving throughput per Joule.
		}
		\label{fig:dl-tpe}
	\end{minipage}
	\caption{Deep learning performance on \soccluster and the traditional edge server.
	\textmd{Background colors are used to distinguish the performance bars/legends of different hardware: 
	\soccluster (white), Intel CPU (grey), NVIDIA GPUs (light blue).
	}
	}
	\vspace{-10pt}
\end{figure*}

\subsection{Transcoding Quality}\label{sec:video-psnr} 
Transcoding quality refers to the consumer perception of the output video.
The live streaming transcoding experiments were conducted with a fixed target bitrate.
However, due to the nuances at both the software (libx264 vs. MediaCodec vs. NVENC) and hardware (CPU vs. GPU vs. ASIC) layers, video quality could vary noticeably even under identical bitrate constraints.
To evaluate this, we saved the output of live streaming transcoding to files from previous experiments and used Peak Signal-to-Noise (PSNR)~\cite{psnr} as the metric to represent video quality, where higher values indicate better quality.

As shown in Figure~\ref{fig:video-enc-quality},
\textit{the software encoder running on SoC CPUs can maintain nearly equivalent video quality to those using the Intel CPU and the NVIDIA GPU, while videos generated by \soccluster's hardware codec have slightly poorer quality than others.}
The general-purpose computing units (i.e., the SoC CPU and the Intel CPU), paired with the software encoder (i.e., libx264) using identical transcoding configurations, always generate videos with the same quality.
In contrast, videos transcoded by MediaCodec exhibit about 1.35\%--14.77\% lower PSNR values compared to those generated by libx264 using SoC CPUs.
This is caused by the less stringent quality and bitrate specifications of mobile encoders~\cite{DBLP:conf/asplos/RanganathanSCDG21}.
To further explore what bitrate should be set to attain comparable video quality using MediaCodec, we manually tuned the target bitrate and then compared the PSNR value with the original video.
Despite these adjustments, videos generated using MediaCodec failed to match the video quality achieved by libx264.
As such, if the quality loss incurred by MediaCodec is not bearable, it is better to choose SoC CPUs for video transcoding.

\subsection{Network Bound Analysis}
\label{sec:video-network}
If the \soccluster is fully occupied with video transcoding workloads, will the network capacity become the bottleneck?
Our mathematical analysis in Table~\ref{tab:transcoding-task} indicates that it will not.
By multiplying the theoretical maximum number of live video streams supported by a single SoC (SoC CPU + SoC hardware codec) and the total network traffic per stream,
we find that among the six videos tested, network usage will slightly exceed the PCB's 1 Gbps capacity only when transcoding V5 video.
In practice, this is not feasible as software delegation daemon processes of SoC hardware codecs also consume some CPU resources.
For the entire \soccluster, the ESB's 20 Gbps capacity will not become a bottleneck.
However, with denser SoC integration in future \soccluster generations or more complex video transcoding workloads, the network could become a limiting factor and need enhancement.

\summary{
\textbf{\textit{Summary.}}
SoC CPUs in \soccluster provide up to 3.2$\times$ and 4.5$\times$ higher energy efficiency than the traditional Intel CPU and NVIDIA GPU, respectively.
The fully-fledged software stack enables a seamless transition of current transcoding services to \socclusters.
Although hardware codecs of SoCs provide even higher throughput and energy efficiency than SoC CPUs, %
their differences in software-level encoding library implementation lead to inconsistent qualities and bitrates in output videos.
Nonetheless, SoC CPUs demonstrate high and robust potential for general video transcoding services, and hardware codecs provide even higher profits for relatively narrower video transcoding scenarios.
}

	\section{Deep Learning Serving Results}\label{sec:dl-serving}

\subsection{Inference Latency}\label{sec:dl-latency}
Inference latency is a crucial factor in delay-sensitive workloads, directly impacting the user experience.
In our study, we conducted tests on NVIDIA GPUs using varying batch sizes to strike a balance between inference latency and energy efficiency.
We limited the batch size to 1 on other hardware, as this setup fully utilizes hardware resources;
increasing the batch size further only resulted in higher latency while not improving energy efficiency.
The results are shown in Figure~\ref{fig:dl-latency}.

We have made the following observations.
(1) The SoC GPUs in \soccluster exhibit 1.55$\times$--2.61$\times$ lower latency compared to SoC CPUs across tasks, and their performance are comparable to the Intel CPU with eight cores used in our experiments.
(2) The latency demonstrated by the NVIDIA GPU with batch size 1 is significantly lower than other processors due to its high hardware-level parallelism and well-optimized software (i.e., TensorRT).
However, when larger batch sizes are used, the latency on the NVIDIA GPU increases significantly and may even approach or exceed that of \soccluster, e.g., performing inference on YOLOv5x (FP32) and BERT (FP32) using the NVIDIA A40 GPU with a batch size of 64.
(3) For medium-sized DNNs, such as ResNet-50, SoC GPUs (for FP32 format) or DSPs (for INT8 format) are typically capable of delivering satisfactory inference latency, i.e., 32.7 ms and 8.8 ms, respectively.
In comparison, using the NVIDIA GPU for inference only results in marginal speed up (e.g., approximately 8 ms for a INT8-based ResNet-50 model), but comes at the cost of higher energy consumption, as we will discuss below.
(4) For large DNNs like ResNet-152, considering both FP32 and INT8 formats, the inference latency of \soccluster ranges from 20.4 ms to 269 ms, which is unacceptable for real-time applications~\cite{arvr-latency}.
To address this issue, a potential solution could be a cooperative inference framework among multiple SoCs.

\begin{figure}[t]
	\centering
	\begin{minipage}[]{0.46\textwidth}
		\includegraphics[width=1\textwidth]{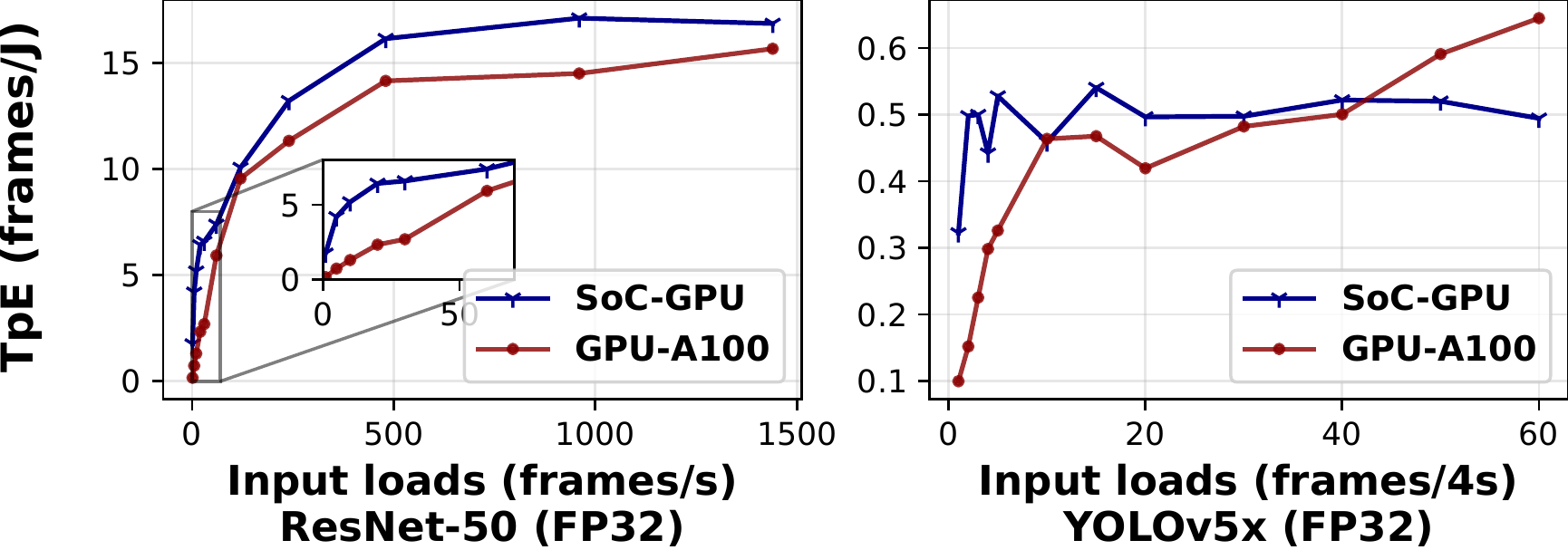}
		\caption{Energy efficiency of \soccluster and the traditional edge server under various DL input loads.
		\textmd{\soccluster delivers higher throughput per energy with light workloads.}
		}
		\label{fig:dl-energy-scale}
		\vspace{-10pt}
	\end{minipage}	
\end{figure}

\subsection{Energy Efficiency}\label{sec:dl-energy}
We use throughput per energy unit as the metric to evaluate the energy efficiency of each hardware type, which is calculated as the number of samples processed per Joule.
A server with higher energy efficiency can process more samples with the same amount of electricity.
The results are presented in Figure~\ref{fig:dl-tpe}.
Our key observation is that \textit{\soccluster exhibits significantly higher energy efficiency compared to the Intel CPU and mid-end NVIDIA GPU, and is comparable to the high-end NVIDIA GPU.}
Especially on ResNet-50 (FP32), SoC GPUs show the ability to process about 18 frames per second per Joule, which is 7.09$\times$ higher than the Intel CPU, 1.78$\times$ higher than the NVIDIA A40 (BS=64), and 1.15$\times$ higher than the NVIDIA A100 (BS=64).
The energy efficiency advantage of \soccluster is even more significant in quantized models.
Taking ResNet-152 with INT8 format as an example, the energy efficiency of SoC DSPs is 42$\times$ higher than that of the Intel CPU and 1.5$\times$ higher than that of the NVIDIA A100 (BS=64).
This can be attributed to the fact that SoC DSPs are designed for low-power data processing, operating at frequencies of $\le$500MHz.

The above energy efficiency values are primarily based on the measurements taken when the servers are fully loaded.
We also measured energy efficiency under varying workloads, as discussed in $\S$\ref{sec:video-energy}.
Figure~\ref{fig:dl-energy-scale} illustrates this analysis of the SoC GPU and the NVIDIA A100 GPU, chosen due to their high energy efficiency on \socclusters and traditional edge servers, respectively.
For ResNet-50, \soccluster shows significant energy efficiency advantages when the workload is lightweight, e.g., 5.71$\times$ more energy-efficient than the NVIDIA A100 GPU on average with only five samples per second.
This is primarily due to \soccluster provides fine-grained scheduling at the level of each SoC for efficient processing of incoming requests.
When incoming data can be adequately processed by only a subset of SoCs, the remaining SoCs can be kept in a low-power state or even turned off.
In comparison, datacenter-level GPUs have coarser granularity to scale their energy consumption with dynamic workloads.

\subsection{SoC-collaborative DL Inference}
\label{sec:dl-collaborative}
In this section, we conducted a preliminary analysis on SoC-collaborative DL inference, which aims to mitigate high inference latency.
We used the MNN~\cite{mnn} framework and the tensor parallelism algorithm proposed in~\cite{zeng2020coedge}.
Specifically, each participating SoC calculates 1/N of the entire tensor along the width dimension.
Intermediate results are transmitted between SoCs via the TCP protocol.

\begin{figure}[t]
	\includegraphics[width=0.45\textwidth]{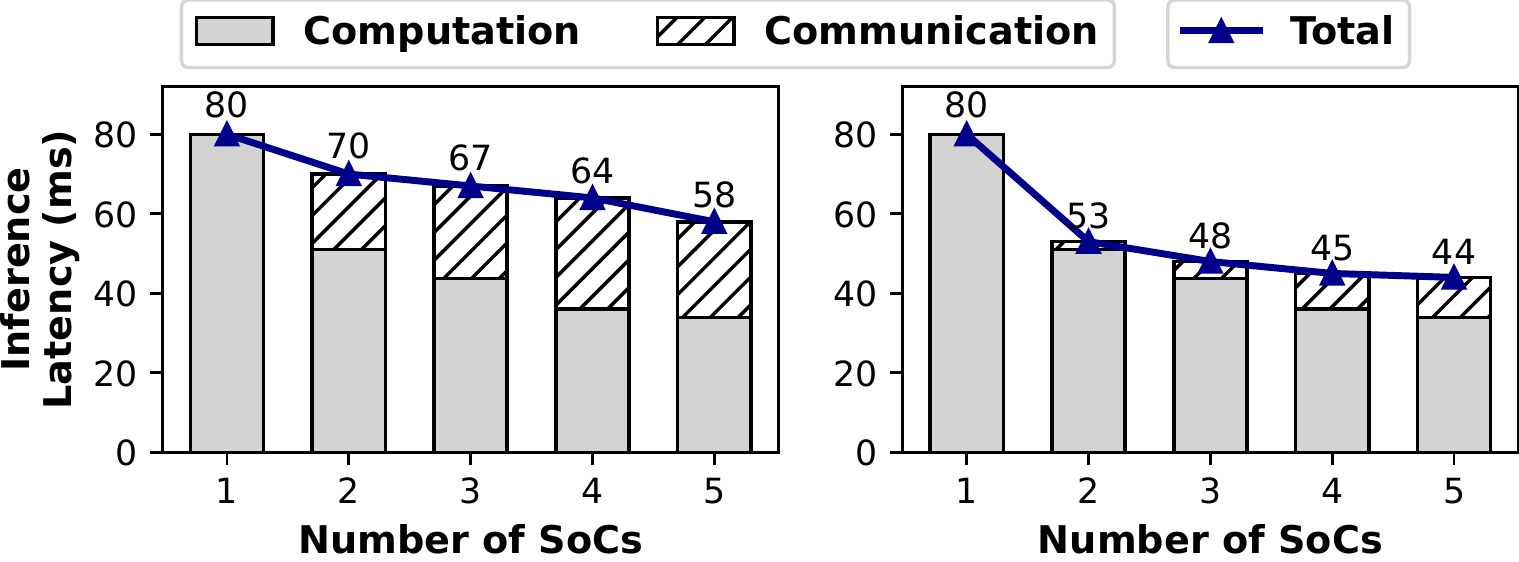}
	\caption{SoC-collaborative DL inference latency using different number of SoCs and breakdown of latencies.
	Left: Tensor parallelism; Right: Tensor parallelism with computation/communication pipelining.}
	\label{fig:net-crossdl}
\end{figure}

Figure~\ref{fig:net-crossdl} shows inference latencies and their breakdown when using 1--5 SoCs.
We observe that involving more SoCs does not proportionally reduce inference latencies.
This may stem from imperfections in the software design that incur additional computation and high communication overhead.
For example, on the ResNet-50 model, increasing the number of SoCs from one to five reduces the computation time from 80 ms to 34 ms (a 2.35$\times$ reduction), while the inference only achieves a 1.38$\times$ speedup.
When using five SoCs, the data communication time contributes to 41.5\% of the total inference latency, where both computation and communication time lengthen the overall latency.
We then tried to optimize the data synchronization design between SoCs by transferring computation-required data first, aiming to pipeline computation and communication.
Results show that the network communication time still accounts for 22.9\% of the total latency with five SoCs involved, indicating network bandwidth could bottleneck the SoC collaboration process.
Therefore, software optimizations (e.g., more fine-grained tensor partitioning) and hardware enhancements (e.g., increased network bandwidth) should be jointly utilized to improve performance, especially when a larger number of SoCs are involved.

\summary{

\textbf{\textit{Summary.}}
SoC GPUs and DSPs in \soccluster demonstrate superior energy efficiency compared to traditional server-level CPUs (up to 42$\times$) and GPUs (up to 6.5$\times$).
The inference latency of \soccluster on medium-sized DNNs, such as ResNet-50, is satisfactory for meeting the requirements of typical edge applications.
However, more advanced software that can orchestrate multiple SoCs is urgently demanded to collaboratively serve large DNNs on \socclusters efficiently.
\soccluster could also enhance its network bandwidth to reduce data communication time in cross-SoC DL inference.

}

	\begin{table*}[h]
    \scalebox{1}{
    \footnotesize
    \begin{tabular}{|l||l|l|l||l|l|}
    \hline
    \textbf{TCO Component}                                                                                 & \textbf{Parameter}                              & \textbf{\begin{tabular}[c]{@{}l@{}}Edge Server\\ Cost\end{tabular}} & \textbf{\begin{tabular}[c]{@{}l@{}}Edge Server\\ (W/O GPU) Cost\end{tabular}} & \textbf{Parameter}           & \textbf{\begin{tabular}[c]{@{}l@{}}SoC Cluster\\ Cost\end{tabular}} \\ \hhline{=#=|=|=#=|=}
    \multirow{7}{*}{\begin{tabular}[c]{@{}l@{}}Capital\\ Expenditure\\ (CapEx)\end{tabular}}               & Intel CPU                                       & \$2,740 (5.7\%)                                                                 & \$2,740 (21.0\%)                                                                          & 60$\times$ SoC                                  & \$24,489 (67.5\%)                                                          \\
                                                                                                           & DRAM                                            & \$3,540 (7.3\%)                                                                 & \$3,540 (27.1\%)                                                                          & 12$\times$ PCB                                  & \$7,075 (19.5\%)                                                           \\
                                                                                                           & Disk                                            & \$1,220 (2.5\%)                                                                 & \$1,220 (9.4\%)                                                                           & Ethernet Switch Board                           & \$689 (1.9\%)                                                              \\ 
                                                                                                           & 8$\times$ NVIDIA A40 GPU                        & \$35,192 (73.0\%)                                                               & \$0 (0\%)                                                                                 & BMC                                             & \$1,923 (5.3\%)                                                            \\ 
                                                                                                           & Others                                          & \$5,544 (11.5\%)                                                                & \$5,544 (42.5\%)                                                                          & Others                                          & \$2,104 (5.8\%)                                                            \\ \cline{2-6} 
                                                                                                           & Total CapEx                                     & \$48,236                                                                        & \$13,044                                                                                  & Total CapEx                                     & \$36,280                                                                   \\
                                                                                                           & Total CapEx/36 months                           & \$1,340                                                                         & \$363                                                                                     & Total CapEx/36 months                           & \$1,008                                                                    \\ \hhline{=#=|=|=#=|=}
    \multirow{5}{*}{\begin{tabular}[c]{@{}l@{}}Operational\\ Expenditure\\ (OpEx)\end{tabular}}            & Avg. peak power consumption                     & 1,231 watts                                                                     & 633 watts                                                                                 & Avg. peak power consumption                     & 589 watts                                                                  \\
                                                                                                           & Monthly kWh (50\% Util.)                        & 443 kWh                                                                         & 228 kWh                                                                                   & Monthly kWh (50\% Util.)                        & 212 kWh                                                                    \\
                                                                                                           & Electricity unit cost~\cite{elec-price}         & \$0.0786/kWh                                                                    & \$0.0786/kWh                                                                              & Electricity unit cost~\cite{elec-price}         & \$0.0786/kWh                                                               \\
                                                                                                           & Server electricity cost                         & \$35                                                                            & \$18                                                                                      & Server electricity cost                         & \$17                                                                       \\
                                                                                                           & PUE Overhead (PUE=2.0~\cite{google-dccomputer}) & \$35                                                                            & \$18                                                                                      & PUE Overhead (PUE=2.0~\cite{google-dccomputer}) & \$17                                                                       \\ \cline{2-6}
                                                                                                           & Monthly electricity cost                        & \$70                                                                            & \$36                                                                                      & Monthly electricity cost                        & \$34                                                                       \\ \hhline{=#=|=|=#=|=}
    Total                                                                                                  & Monthly TCO                                     & \$1,410                                                                         & \$399                                                                                     & Monthly TCO                                     & \$1,042                                                                    \\ \hline
    \end{tabular}

    }
    \caption{Capital expenditure (CapEx), operational expenditure (OpEx), and resultant monthly TCO of each server.
    \textmd{We additionally estimated the TCO of the traditional edge server without NVIDIA GPUs.
    The monthly electricity cost was calculated by sampling the average peak power consumption when performing live streaming transcoding on V5.
    }
    }
    \vspace{-10pt}
    \label{tab:tco}
    \end{table*}

\section{Cost Analysis}\label{sec:tco-analysis}
Cost is another critical dimension to evaluate the suitability of new hardware.
In this section, we conduct a TCO analysis on \soccluster and traditional edge servers.
The TCO consists of two parts: the capital expenditure needed to purchase the servers (CapEx, with a breakdown of each hardware component) and the operational expenditure (OpEx).
We calculate the total CapEx using the retail purchase cost.
For OpEx, we consider only the electricity cost as in prior work~\cite{lang2010wimpy}.
In addition to the traditional edge server that consists of an Intel CPU and 8 NVIDIA GPUs, we add a ``virtual server'' by excluding all 8 NVIDIA GPUs.
We use throughput per cost (TpC) as the normalized performance metric.
A server with a higher TpC value indicates its ability to process more workloads within the same monetary cost budget.

\noindent \textbf{Capital expenditure.}
As shown in the top half part of Table~\ref{tab:tco}, GPUs account for a substantial portion of the total CapEx in a traditional edge server.
In a CPU-only server, the CapEx is amortized across different hardware components.
Within \soccluster, 60 SoCs and 12 PCBs constitute almost 87\% of the total CapEx.
In summary, \soccluster has a lower CapEx than the traditional edge server with 8 NVIDIA GPUs but costs about 2.8$\times$ more than a CPU-only edge server.

\noindent \textbf{Operational expenditure.}
We opted not to use complex OpEx models~\cite{shahrad2017towards} as we observed that CapEx consistently dominated the TCO, as we describe later.
This was also reported in a previous study by Google~\cite{google-dccomputer}.
Therefore, we have chosen to report only the electricity cost in this analysis.
The monthly electricity cost is calculated by multiplying the monthly power consumption (kWh) with the electricity unit cost (\$/kWh).
The monthly power consumption across all workloads correlates with their average power usage.
For instance, performing live streaming transcoding when fully utilizing all 8 NVIDIA A40 GPUs has an average power consumption of 1,231 watts.
Assuming the servers operate at their average peak power 50\% of the time over a month, the monthly power consumption can be calculated as $1231W*50\%*24h*30/1000=443kWh$.
The electricity unit cost was ascertained by referring to the U.S. industrial average electricity price over one year, from August 2021 to July 2022~\cite{elec-price}.
Thus, the monthly electricity cost directly related to computation is $\$0.0786/kWh*443kWh\approx\$35$.
Additionally, the PUE (Power Usage Effectiveness) overhead, reflected by the ratio of total building power consumption to IT infrastructure power consumption, adds to the monthly electricity cost~\cite{google-dccomputer}.
We used a slightly higher PUE value (2.0) at the edge, compared to 1.5 at cloud data centers~\cite{google-dccomputer}.
The overall monthly electricity cost is $\$35+\$35*(2.0-1)=\$70$.

\noindent \textbf{Monthly cost.}
The detailed monthly cost calculation is presented in Table~\ref{tab:tco}.
In line with prior work~\cite{lang2010wimpy}, we break down the monthly TCO into the following components:

\begin{itemize}
    \item \textit{Total CapEx amortized to 36 months.} 
    By assuming a 3-year server lifetime~\cite{lang2010wimpy,janapa2010web,google-dccomputer}, we amortized the CapEx of each server to 36 months.
    \item \textit{Monthly OpEx} mainly refers to the electricity cost.
    It is worth noting that the monthly OpEx is significantly less than the amortized CapEx (e.g., \$70 vs. \$1,340 for the traditional edge server).
\end{itemize}

We added these two expenditure figures to get the monthly TCO, then normalized the application throughput (measured in previous sections) to the monthly TCO as the TpC metric.
Table~\ref{tab:tco-vertical} shows that \soccluster is a cost-efficient option for live streaming transcoding.
Specifically, compared to the edge server with NVIDIA A40 GPUs, the SoC CPUs show a geometric mean of TpC that is 4.28$\times$ higher than the Intel CPU and 2.23$\times$ higher than NVIDIA GPUs.
Even without NVIDIA GPUs in the traditional edge server, SoC CPUs achieve a geometric mean of TpC that is 1.22$\times$ higher than the Intel CPU.
Moreover, the Intel CPU in the non-GPU server shows higher TpC than NVIDIA A40 GPUs for all videos.
For archive transcoding, \soccluster is less cost-efficient than the traditional edge server due to its low throughput on a single SoC and relatively high CapEx.
For most scenarios, processing archive transcoding with the NVIDIA GPUs provides higher TpC compared to other hardware options.
Regarding DL serving, the NVIDIA GPUs exhibit a marked increase in cost efficiency over \socclusters.
This is mainly attributed to their ability to handle batched DL requests and deliver high throughput under full loads.

\begin{table}[t]
	\centering
    \scalebox{0.66}{
    \begin{tabular}{|l|l|llllll|}
    \hline
    \multirow{2}{*}{\textbf{Server}}                                                   & \multirow{2}{*}{\textbf{Hardware}} & \multicolumn{6}{c|}{\textbf{Live Streaming Transcoding TpC (streams/\$)}}                                                                                                                                                                                                                                                                                                                                                                                    \\ \cline{3-8} 
                                                                                       &                                    & \multicolumn{1}{l|}{V1}                                                   & \multicolumn{1}{l|}{V2}                                                    & \multicolumn{1}{l|}{V3}                                                    & \multicolumn{1}{l|}{V4}                                                    & \multicolumn{1}{l|}{V5}                                                   & V6                                                    \\ \hline
    \multirow{2}{*}{\begin{tabular}[c]{@{}l@{}}Edge (W/ GPU)\end{tabular}} & Intel CPU                          & \multicolumn{1}{l|}{0.180}                                                & \multicolumn{1}{l|}{0.223}                                                 & \multicolumn{1}{l|}{0.057}                                                 & \multicolumn{1}{l|}{0.101}                                                 & \multicolumn{1}{l|}{0.042}                                                & 0.013                                                 \\ \cline{2-8} 
                                                                                       & GPU A40                            & \multicolumn{1}{l|}{0.420}                                                & \multicolumn{1}{l|}{0.210}                                                 & \multicolumn{1}{l|}{0.102}                                                 & \multicolumn{1}{l|}{0.181}                                                 & \multicolumn{1}{l|}{0.114}                                                & 0.034                                                 \\ \hline
    \begin{tabular}[c]{@{}l@{}}Edge (W/O GPU)\end{tabular}      & Intel CPU                          & \multicolumn{1}{l|}{0.627}                                                & \multicolumn{1}{l|}{0.777}                                                 & \multicolumn{1}{l|}{0.200}                                                 & \multicolumn{1}{l|}{0.351}                                                 & \multicolumn{1}{l|}{0.146}                                                & 0.047                                                 \\ \hline
    SoC Cluster                                                                        & SoC-CPU                            & \multicolumn{1}{l|}{\cellcolor{pink}{0.748}}                              & \multicolumn{1}{l|}{\cellcolor{pink}{0.863}}                               & \multicolumn{1}{l|}{\cellcolor{pink}{0.230}}                               & \multicolumn{1}{l|}{\cellcolor{pink}{0.519}}                               & \multicolumn{1}{l|}{\cellcolor{pink}{0.173}}                              & \cellcolor{pink}{0.058}                                           \\ \hline \hline
    \multirow{2}{*}{\textbf{Server}}                                                   & \multirow{2}{*}{\textbf{Hardware}} & \multicolumn{6}{c|}{\textbf{Archive Transcoding TpC (frames/s/\$)}}                                                                                                                                                                                                                                                                                                                                                                                   \\ \cline{3-8} 
                                                                                       &                                    & \multicolumn{1}{l|}{V1}                                                   & \multicolumn{1}{l|}{V2}                                                    & \multicolumn{1}{l|}{V3}                                                    & \multicolumn{1}{l|}{V4}                                                    & \multicolumn{1}{l|}{V5}                                                   & V6                                                    \\ \hline
    \multirow{2}{*}{\begin{tabular}[c]{@{}l@{}}Edge (W/ GPU)\end{tabular}} & Intel CPU                          & \multicolumn{1}{l|}{0.027}                                                & \multicolumn{1}{l|}{0.053}                                                 & \multicolumn{1}{l|}{0.020}                                                 & \multicolumn{1}{l|}{0.024}                                                 & \multicolumn{1}{l|}{0.004}                                                & 0.001                                                 \\ \cline{2-8} 
                                                                                       & GPU A40                            & \multicolumn{1}{l|}{\cellcolor{pink}{0.162}}                              & \multicolumn{1}{l|}{0.140}                                                 & \multicolumn{1}{l|}{\cellcolor{pink}{0.203}}                               & \multicolumn{1}{l|}{\cellcolor{pink}{0.086}}                               & \multicolumn{1}{l|}{\cellcolor{pink}{0.091}}                              & \cellcolor{pink}{0.035}                                           \\ \hline
    \begin{tabular}[c]{@{}l@{}}Edge (W/O GPU)\end{tabular}      & Intel CPU                          & \multicolumn{1}{l|}{0.094}                                                & \multicolumn{1}{l|}{{\cellcolor{pink}0.189}}                               & \multicolumn{1}{l|}{0.072}                                                 & \multicolumn{1}{l|}{0.085}                                                 & \multicolumn{1}{l|}{0.013}                                                & 0.004                                                 \\ \hline
    SoC Cluster                                                                        & SoC-CPU                            & \multicolumn{1}{l|}{0.015}                                                & \multicolumn{1}{l|}{0.046}                                                 & \multicolumn{1}{l|}{0.010}                                                 & \multicolumn{1}{l|}{0.022}                                                 & \multicolumn{1}{l|}{0.002}                                                & \textless{}0.001                                      \\ \hline \hline
    \multirow{2}{*}{\textbf{Server}}                                                   & \multirow{2}{*}{\textbf{Hardware}} & \multicolumn{6}{c|}{\textbf{DL Serving TpC (frames/s/\$)}}                                                                                                                                                                                                                                                                                                                                                                                            \\ \cline{3-8} 
                                                                                       &                                    & \multicolumn{1}{l|}{\begin{tabular}[c]{@{}l@{}}R-50\\ (FP32)\end{tabular}} & \multicolumn{1}{l|}{\begin{tabular}[c]{@{}l@{}}R-152\\ (FP32)\end{tabular}} & \multicolumn{1}{l|}{\begin{tabular}[c]{@{}l@{}}YOLO\\ (FP32)\end{tabular}} & \multicolumn{1}{l|}{\begin{tabular}[c]{@{}l@{}}BERT\\ (FP32)\end{tabular}} & \multicolumn{1}{l|}{\begin{tabular}[c]{@{}l@{}}R-50\\ (INT8)\end{tabular}} & \begin{tabular}[c]{@{}l@{}}R-152\\ (INT8)\end{tabular} \\ \hline
    \multirow{2}{*}{\begin{tabular}[c]{@{}l@{}}Edge (W/ GPU)\end{tabular}} & Intel CPU                          & \multicolumn{1}{l|}{0.579}                                                 & \multicolumn{1}{l|}{0.176}                                                  & \multicolumn{1}{l|}{0.010}                                                 & \multicolumn{1}{l|}{0.044}                                                 & \multicolumn{1}{l|}{1.201}                                                 & 0.355                                                 \\ \cline{2-8} 
                                                                                       & GPU A40                            & \multicolumn{1}{l|}{\cellcolor{pink}{14.631}}                              & \multicolumn{1}{l|}{\cellcolor{pink}{4.535}}                                & \multicolumn{1}{l|}{\cellcolor{pink}{0.571}}                               & \multicolumn{1}{l|}{\cellcolor{pink}{7.311}}                               & \multicolumn{1}{l|}{\cellcolor{pink}{45.684}}                              & \cellcolor{pink}{19.840}                                          \\ \hline
    \begin{tabular}[c]{@{}l@{}}Edge (W/O GPU)\end{tabular}      & Intel CPU                          & \multicolumn{1}{l|}{2.026}                                                 & \multicolumn{1}{l|}{0.617}                                                  & \multicolumn{1}{l|}{0.036}                                                 & \multicolumn{1}{l|}{0.152}                                                 & \multicolumn{1}{l|}{4.199}                                                 & 1.242                                                 \\ \hline
    \multirow{3}{*}{SoC Cluster}                                                       & SoC-CPU                            & \multicolumn{1}{l|}{0.750}                                                 & \multicolumn{1}{l|}{0.131}                                                  & \multicolumn{1}{l|}{0.026}                                                 & \multicolumn{1}{l|}{1.840}                                                 & \multicolumn{2}{c|}{-}                                                                                                            \\ \cline{2-8} 
                                                                                       & SoC-GPU                            & \multicolumn{1}{l|}{3.210}                                                 & \multicolumn{1}{l|}{0.628}                                                  & \multicolumn{1}{l|}{0.077}                                                 & \multicolumn{3}{c|}{-}                                                                                                                                                                                         \\ \cline{2-8} 
                                                                                       & SoC-DSP                            & \multicolumn{4}{c|}{-}                                                                                                                                                                                                                                                                                           & \multicolumn{1}{l|}{6.673}                                                & 2.871                                                 \\ \hline
    \end{tabular}
    }
    \caption{Normalized application throughput to monthly TCO.
    \textmd{We highlight the highest throughput per cost among used hardware for each video/model.
    ``Edge'': the traditional edge server.
    }
    \vspace{-10pt}
    }
    \label{tab:tco-vertical}
\end{table}

\summary{
\textbf{\textit{Summary.}}
\soccluster's CapEx is comparable to an 8-GPU server and is significantly higher than that of a CPU server.
While \soccluster has the potential to reduce OpEx through electricity savings, the total cost is still dominated by the CapEx.
Regarding specific workloads, \soccluster outperforms traditional CPU/GPU servers in cost efficiency for live streaming transcoding.
However, it falls behind the NVIDIA GPU for tasks that demand significant computing capacity (archive transcoding), and those involving highly GPU-optimized workloads (DL serving).
}

	\vspace{-5pt}
\section{SoC Longitudinal Study}\label{sec:soc-evol}
\vspace{-5pt}

\begin{figure}[t]
	\centering

	\includegraphics[width=0.44\textwidth]{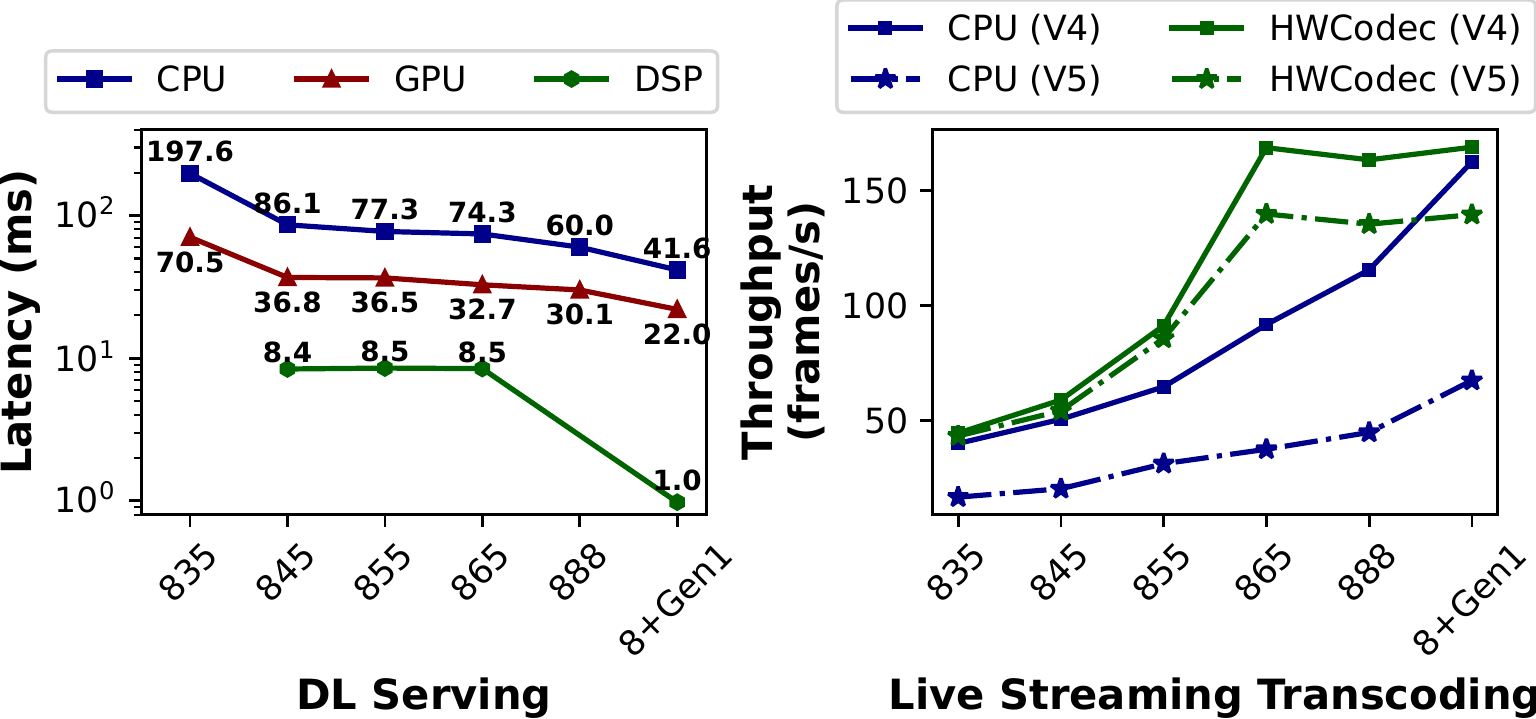}
	\vspace{-5pt}
	\caption{Performance evolution of six high-end Qualcomm Snapdragon SoCs released between 2017--2022.}
	\vspace{-15pt}
	\label{fig:soc-evol}
\end{figure}

The previous experiments were conducted on an \soccluster consisting of a specific SoC model, i.e., the Qualcomm Snapdragon 865 released in 2020.
To expand the observations to more hardware types and, more importantly, to understand SoC performance evolution over time, we performed a longitudinal study on six mobile SoCs released from 2017 to 2022.
We selected six smartphones equipped with high-end Qualcomm Snapdragon SoCs, with their specifications listed in Table~\ref{tab:soc-models}.
We repeated our experiments on two workloads and presented the results in Figure~\ref{fig:soc-evol}.
\begin{table}[t]
    \footnotesize
    \centering
    \scalebox{0.9}{
    \begin{tabular}{|l|l|l|l|l|}
    \hline
    \textbf{Devices} & \textbf{SoC} & \textbf{RAM} & \textbf{OS}   & \textbf{Release Date}  \\ \hline
    Xiaomi 12 S      & QS 8+Gen1  & 12 GB        & Android 12    & May 2022        \\ \hline
    Xiaomi 11 Pro    & QS 888       & 8 GB         & Android 11    & Jun. 2021       \\ \hline
    Meizu 17         & QS 865       & 8 GB         & Android 10    & Mar. 2020       \\ \hline
    Meizu 16T        & QS 855       & 6 GB         & Android 9     & Mar. 2019       \\ \hline
    Xiaomi 8         & QS 845       & 6 GB         & Android 8.1   & Feb. 2018       \\ \hline
    Xiaomi 6         & QS 835       & 6 GB         & Android 7.1.1 & Mar. 2017       \\ \hline
    \end{tabular}
    }
    \vspace{-5pt}
    \caption{Device specifications of six mobile phones used in the SoC longitudinal study.
    \textmd{QS: Qualcomm Snapdragon.
    }
    }
    \vspace{-13pt}
    \label{tab:soc-models}
\end{table}

First, we measured the DL serving latency on ResNet-50 using the same experimental settings as in $\S$\ref{sec:dl-serving}.
The results reveal a significant performance boost in SoC DSPs, with an 8.4$\times$ inference latency reduction from the Snapdragon 845 (2018) to the Snapdragon 8+Gen1 (2022).
SoC CPUs and GPUs also show improvements, but not as significantly as DSPs, with latency reductions of 4.8$\times$ and 3.2$\times$ from 2017 to 2022, respectively.
An additional experiment focusing on inference throughput shows that the latest Snapdragon 8+Gen1 phone achieved 1.7$\times$ higher throughput on its DSP when setting the batch size to 8, compared to the default setup employing a batch size of 1.
Furthermore, the recent incorporation of support for floating-point calculations on Qualcomm's flagship Hexagon DSPs~\cite{8gen2} has positioned these processors as suitable candidates to serve increasingly complex AI tasks~\cite{qcom-sdmobile,touvron2023llama,xu2022mandheling}.

In live streaming transcoding experiments, we measured frames processed per second during the transcoding of two fixed-duration videos (V4 and V5, with metadata detailed in Table~\ref{tab:transcoding-task}).
Our key observation is that live streaming transcoding tasks reflect a pattern of gradual performance improvement similar to that seen in DL serving experiments.
When using SoC CPUs, the throughput for video V4 on the Snapdragon 865 is 1.42$\times$, 1.82$\times$, and 2.3$\times$ higher than that on 855, 845, and 835, respectively.
Furthermore, this value increased by 1.8$\times$ on the 8+Gen1 phone.
This trend is even more impressive on the hardware codec -- the throughput on the Snapdragon 865 was 3.8$\times$ and 3.24$\times$ greater than that on the 835 for V4 and V5, respectively.
These performance improvements are attributed not only to the evolution of hardware but also to the enhancements of software along with the Android OS upgrades~\cite{android11-dec,android12-enc}.

To conclude, mobile SoCs have shown tremendous performance improvements in the past six years, positioning them as promising candidates for more complex server-side workloads.
As Moore's Law approaches its limits, CPU performance evolution might slow down, but the capability of mobile co-processors (GPUs, DSPs, and hardware codecs) continue to grow at a rapid pace, and their performance gains over CPU are also on the rise.
To achieve a sustained performance evolution for mobile SoC-based servers, it is pivotal to strategically leverage these co-processors.

	\vspace{-10pt}
\section{Discussion and Future Work}\label{sec:discussion}
\vspace{-5pt}

\noindent \textbf{Applicability of measurement results.}
The primary goal of this study is to inspire the edge computing community with a new form of edge server, paving the way towards more efficient computing under constrained deployment environments (space, electricity, etc.).
Currently, the measurement results are conducted on a specific \soccluster with publicly available commercial services for user access~\cite{ecp}.
However, we believe our primary observations could apply to a broader range of \soccluster-like hardware models and benefit the community further.
First, in terms of performance, the longitudinal study on different SoC models over years ($\S$\ref{sec:soc-evol}) may indicate improved performance of \socclusters equipped with advanced SoC models.
Second, improving the networking subsystem may significantly speed up data-intensive applications, such as collaborative DL training.
The high energy efficiency of mobile SoCs usually makes them good candidates for workloads they can properly serve at the edge.

\noindent \textbf{Overhead of SoC virtualization.}
The primary results on Intel CPUs in this study are measured within Docker containers, with the same experiments conducted on physical SoCs without virtualization.
For a fair comparison, we measure DL serving application performance and hardware resource usage on both physical and virtualized SoCs.
Currently, the virtualization solution of \soccluster allows a physical SoC to only run the Android Linux kernel.
A virtualized SoC uses the same Linux OS and is required to run the Android framework inside Docker containers, which leads to higher memory usage as shown in Table~\ref{tab:dl-docker-overhead}. 
In most cases, there are trivial differences in DL serving performance and hardware resource usage.
One exception is that the immature Android containerization solution prevents GPU workloads on virtualized SoCs from achieving the same high level of GPU usage as on physical SoCs, leading to a slight performance slowdown (e.g., 60 ms on the YOLOv5x model).
We believe the SoC virtualization implementation can be further improved to mitigate the potential performance overhead and unnecessary resource usage when the Android framework is not required.

\setlength{\tabcolsep}{2pt} %
\begin{table}[t]
    \centering
    \scalebox{0.67}{

    \begin{tabular}{|c|c|cc|cc|cc|}
    \hline
    {}                                      & {}                                  & \multicolumn{2}{c|}{{\textbf{SoC CPU}}}         & \multicolumn{2}{c|}{{\textbf{SoC GPU}}}         & \multicolumn{2}{c|}{{\textbf{SoC DSP}}}         \\ \cline{3-8} 
    \multirow{-2}{*}{{\textbf{Model}}}      & \multirow{-2}{*}{{\textbf{Metric}}} & {\textbf{Phy.}} & {\textbf{Vir.}} & {\textbf{Phy.}} & {\textbf{Vir.}} & {\textbf{Phy.}} & {\textbf{Vir.}} \\ \hline
    {}                                      & {Latency}                      & {$81.2/0.2$}          & {$80.4/0.1$}          & {$32.5/0.4$}          & {$33.9/0.1$}          & {$11.0/0.1$}           & {$10.5/0.01$}           \\
    {}                                      & {CPU}                          & {$52.1/0.3$}          & {$53.1/0.1$}          & {$9.0/5.7$}           & {$9.5/0.1$}           & {$5.2/2.1$}           & {$5.7/0.3$}           \\
    {}                                      & {GPU}                          & {$0.7/0.3$}           & {$0.5/0.1$}           & {$73.9/1.2$}          & {$71.3/0.6$}          & {$0.6/0.1$}           & {$0.6/0.1$}           \\
    \multirow{-4}{*}{{\textbf{R50}}}        & {Mem}                          & {$32.3/0.7$}          & {$37.7/0.2$}          & {$35.2/5.7$}          & {$37.6/0.3$}          & {$32.7/1.9$}          & {$37.4/0.2$}          \\ \hline
    {}                                      & {Latency}                      & {$258.3/0.4$}         & {$257.8/1.0$}         & {$100.9/0.1$}         & {$102.8/0.2$}         & {$21.0/0.04$}          & {$20.4/0.02$}          \\
    {}                                      & {CPU}                          & {$53.3/0.1$}          & {$53.9/0.3$}          & {$5.9/0.5$}           & {$8.6/0.1$}           & {$6.0/0.9$}           & {$7.1/0.5$}           \\
    {}                                      & {GPU}                          & {$0.4/0.1$}           & {$0.6/0.1$}           & {$81.1/0.5$}          & {$78.5/0.2$}          & {$0.7/0.1$}           & {$0.6/0.1$}           \\
    \multirow{-4}{*}{{\textbf{R152}}}       & {Mem}                          & {$34.9/1.0$}          & {$40.1/0.1$}          & {$35.6/2.1$}          & {$39.8/0.1$}          & {$33.7/0.8$}          & {$39.0/0.2$}          \\ \hline
    {}                                      & {Latency}                      & {$1121.3/13.7$}        & {$1113.9/2.8$}       & {$620.6/1.0$}         & {$683.7/4.1$}         & \multicolumn{2}{c|}{{}}                                \\
    {}                                      & {CPU}                          & {$53.9/0.2$}          & {$54.5/0.1$}          & {$5.3/0.2$}           & {$7.6/0.1$}           & \multicolumn{2}{c|}{{}}                                \\
    {}                                      & {GPU}                          & {$0.5/0.1$}           & {$0.6/0.04$}          & {$82.5/0.1$}          & {$77.1/0.4$}          & \multicolumn{2}{c|}{{}}                                \\
    \multirow{-4}{*}{{\textbf{YOLO}}}       & {Mem}                          & {$40.1/0.6$}          & {$45.9/0.1$}          & {$39.5/3.3$}          & {$44.2/0.4$}          & \multicolumn{2}{c|}{\multirow{-4}{*}{{/}}}             \\ \hline
    \end{tabular}
    }
    \footnotesize
    \vspace{-5pt}
    \caption{DL inference performance and hardware usages (average/standard deviation) on physical SoCs (Phy.) and virtualized SoCs (Vir.).
        Latency is measured in milliseconds.
        CPU, GPU, and memory utilization are represented as percentages.
        }
    \label{tab:dl-docker-overhead}
    \vspace{-15pt}
    \end{table}

There are also a few directions to further explore to enhance the vision of this study.

\noindent $\bullet$ \textit{Killer applications}. 
\soccluster is endowed with ample storage space and high I/O speed, making it well-suited for database systems with compatible design patterns~\cite{mobisys19-flash}.
The SoC-level workload scheduling granularity lends itself to ephemeral serverless workloads~\cite{atc20-serverless-trace}.
However, mobile SoCs are not typically designed to operate at full speed and 24/7 in clouds, presents a challenge for operating \soccluster.
The failure of a single SoC subsystem, such as flash, can render the application and entire SoC unusable. 
Therefore, fault tolerance is crucial for the success of \soccluster.

\noindent $\bullet$ \textit{Operating system}.
Mobile OSs are designed and optimized for interactive scenarios rather than server workloads.
Although it is viable to run Linux or Windows~\cite{snapdragon-x} on ARM SoCs, simply replacing Android with other OSs may result in the loss of compatibility with native mobile apps, as well as the inability to leverage certain hardware accelerators if their drivers are vendor-specific and proprietary~\cite{wu2013impact}.
To get the most from both, the correct approach seems to revise the original Android OS to fit edge workloads.

\noindent $\bullet$ \textit{Network infrastructure and topology.}
We show in $\S$\ref{sec:video-trans} that an overall 20 Gbps network capacity is mostly sufficient to support video transcoding on all SoCs.
However, the limited bandwidth between SoCs inside \socclusters makes the current network infrastructure is not equipped for workloads requiring high-volume data exchanges across SoCs.
High performance datacenter network interfaces and switches, such as InfiniBand~\cite{infiniband} and NVLink~\cite{nvlink}, provide network bandwidth in the hundreds of Gbps, which is two orders of magnitude higher than the 1 Gbps theoretical bandwidth between SoCs in \socclusters.
To support a wider range of data-intensive applications, \soccluster should incorporate recent progress from network research and industry~\cite{infiniband,programmable-switch-asic,sivaraman2020network}.

	\section{Related Work}\label{sec:related}
\vspace{-5pt}

\noindent \textbf{Grouping mobile SoCs as servers.}
Several previous attempts have been made to conceptualize a server consisting of compact SoCs.
Some researchers have investigated whether mobile SoCs can provide sufficient performance and reduce costs for HPC~\cite{rajovic2013supercomputing,rajovic2014tibidabo}.
In an effort to reduce e-waste, Shahrad \textit{et al.}~\cite{shahrad2017towards} constructed computational nodes with used smartphones, but they analyzed server design without evaluating real workloads.
Switzer \textit{et al.} created a junkyard cluster~\cite{junkyardcomputing} comprising just ten smartphones, with an emphasis on reducing carbon footprints.
Our previous study proposed a similar vision of renovating mobile SoCs at the edge~\cite{sec22soc}, but limited experiments and evaluation did not fully reveal the capabilities of \socclusters.
Other studies have employed IoT/mobile SoCs to support specific applications, such as video transcoding~\cite{liu2016greening}, key-value storage~\cite{andersen2009fawn}, web search~\cite{janapa2010web}, and parallel computing~\cite{busching2012droidcluster}.
However, these studies mainly focus on specialized app types and lack a performance comparison with traditional servers.
In contrast, our study utilized a commercial-off-the-shelf \soccluster, conducted application-driven measurements, and presented extensive performance metrics to showcase its suitability for modern, computation-intensive edge workloads.

\noindent \textbf{Energy-efficient cloud/edge.}
Energy efficiency has been recognized as a crucial factor in data centers~\cite{deng2014harnessing}.
Various techniques have been explored to advance green data centers, including workload scheduling and management~\cite{lee2021greendimm,liu2013arbitrating,liu2012renewable,ran2019deepee,zhou2015carbon}, resource under-provisioning~\cite{manousakis2015coolprovision,zhang2021flex}, greening the data-center network~\cite{nsdi10-elastictree,sigcomm12-green,guo2020aggreflow}, among others.
In contrast to these software-level approaches, we propose a redesign of servers to fundamentally enhance energy efficiency.
As the edge infrastructure is still in its preliminary stage, we consider such a radical measure to be feasible.

	\vspace{-13pt}
\section{Conclusion}\label{sec:conclusions}
\vspace{-8pt}

In this study, we explored the feasibility and implications of utilizing a novel type of edge server, \soccluster, which comprises multiple mobile SoCs.
We conducted extensive benchmarking on two typical edge workloads (i.e., video transcoding and DL serving), and quantitatively demonstrated \soccluster achieves substantial energy savings compared to conventional edge servers, while also identifying its limitations.
These findings highlight the promising potential of \soccluster at the edge, and provide guidance for further improvements in both hardware and software.

	\vspace{-10pt}
	\section*{Acknowledgments}
	\vspace{-5pt}
	We sincerely thank our shepherd, Mohammad Shahrad, and the anonymous reviewers for their constructive feedback and efforts, which significantly improved our work.
	This work was supported by National Key R\&D Program of China (No.2021ZD0113001), NSFC (62102045, 62032003, 61921003), and China Institute
	of IoT (Wuxi). Mengwei Xu is the corresponding author of this work. 

	\bibliographystyle{plain}
	\bibliography{ref-mwx}
	
\end{document}